# On the Molecular Picture and Interfacial Temperature Discontinuity During Evaporation and Condensation

Gang Chen*

Department of Mechanical Engineering

Massachusetts Institute of Technology

**Abstract**

Although it has been shown experimentally that a temperature discontinuity exists at the liquid-vapor interface during evaporation and condensation, quantitatively modeling this temperature jump has been difficult. The classical Schrage equation does not give enough information to determine the interfacial temperature jump. Starting from the Boltzmann transport equation, this paper establishes three interfacial boundary conditions to connect the temperature, density, and pressure jumps at the liquid-vapor interface to the interfacial mass and heat fluxes: one for the mass flux (the Schrage equation), one for the heat flux, and the third for the density discontinuities. These expressions can be readily coupled to heat and mass transport equations in the continuum of the liquid and the vapor phases, enabling one to determine the values of the interfacial temperature, density, and pressure jumps. Comparison with past experiments is favorable. A thermomolecular emission model, mimicking thermionic emission of electrons, is also presented to gain more molecular-level insights on the thermal evaporation processes.

* Email: gchen2@mit.edu

## 1. INTRODUCTION

Evaporation and condensation processes happen widely in nature and industrial technologies [1–4]. Interfacial temperature discontinuity in these processes has been recognized [5,6], supported by direct measurements across liquid-vapor interfaces [7–14], kinetic theory-based modeling [15–22], Monte Carlo simulation [23], and molecular dynamics simulations[24–27]. Although some heat transfer models recognized the existence of such temperature discontinuities, most of the time, the temperature discontinuity is neglected. There seems to be ambiguity on the physical mechanisms of the temperature discontinuity at the liquid-vapor interface during phase-change heat transfer, as well as a lack of convenient physics-based models to treat the interfacial discontinuities. The objectives of this paper are to provide simplified kinetic theory-based interfacial conditions for determining the discontinuities in temperature, density, and pressure and a microscopic emission picture for the thermal evaporation process. These interfacial conditions can be combined with continuum equations for modeling phase-change processes.

The literature on evaporation and condensation is vast and cannot possibly be reviewed within this short introduction. Here, we will briefly mention key historical developments. Pioneering studies of evaporation were carried out by Hertz [28] and Knudsen [29]. Hertz measured evaporation of mercury. Knudsen also measured mercury and introduced the concept of the coefficient of evaporation, $\alpha$. Langmuir made original contributions to evaporation through his study of tungsten evaporation at high temperatures [30]. Knacke and Stranski [31] gave a good summary of the experiments and understanding of evaporation mechanisms prior to 1956. The Hertz-Knudsen equation, which can be readily derived from the kinetic theory by neglecting any collision of the molecules, gives the evaporation mass flux between two parallel plates:

$$\dot{m}_{HK} \approx \alpha \sqrt{\frac{m}{2\pi k_B}} \left( \frac{P_s(T_l)}{\sqrt{T_l}} - \frac{P_v}{\sqrt{T_v}} \right) = \alpha \sqrt{\frac{M}{2\pi \mathrm{R}}} \left( \frac{P_s(T_l)}{\sqrt{T_l}} - \frac{P_v}{\sqrt{T_v}} \right), \quad (1)$$

where $P_s$ is the saturation pressure corresponding to the liquid surface temperature $T_l$, and $P_v$ and $T_v$ are the vapor phase pressure and temperature near the interface, $k_B$ the Boltzmann constant, m the mass of a single molecular, R the universal gas constant, M the molar mass, and $\alpha$ the accommodation coefficient.

Schrage [32] in his PhD thesis also gave a good review of the prior work, including work before Hertz. He argued that the Hertz and Knudsen formula did not consider the drift velocity of the evaporating molecules. After considering molecular drift, the equation he arrived at differs from Eq. (1) in the prefactor,

$$\dot{m}_S \approx \frac{2\alpha}{2-\alpha}\sqrt{\frac{M}{2\pi \mathrm{R}}}\left(\frac{P_S(T_l)}{\sqrt{T_l}} - \frac{P_v}{\sqrt{T_v}}\right) \qquad (2)$$

The Schrage equation is widely used as a fundamental relation for evaporative and condensation heat transfer. Knowing the evaporation and condensation rates, one usually finds the heat transfer rate by multiplying $\dot{m}_S$ and the latent heat of evaporation. We show in this article that the latter is at best only an approximation.

Although Schrage derived the molecular distribution function near the interface that is consistent with the drift of the molecules, his theory did not include the effect of heat transfer across the interface. The molecular flux expression Eq. (2) depends on $T_l$, $P_v$ and $T_v$, which are usually not known themselves. Coupling Eq. (2) with continuum descriptions on the liquid and the vapor phases do not give enough equations to determine these unknowns. The exact location of $P_v$ and $T_v$ in Eqs. (1) and (2) is not clear, although it is generally considered to be at the edge of the Knudsen layer, i.e., with several mean free path from the liquid-vapor interface. Furthermore, there are no easy ways to determine these quantities in typical heat transfer situations such as film evaporation or condensation. Based on solutions to the Boltzmann transport equation (BTE), different approximate expressions have been given relating the interfacial mass and heat fluxes to the liquid side pressure ($P_s$) and temperature ($T_l$), the vapor side pressure ($P_v$) and temperature ($T_v$), and the accommodation coefficient ($\alpha$) [18–20,22,33] . It is fair to say that the interfacial heat flux expressions derived by different authors have not come to the same level of clarities as in the Hertz-Knudsen and Schrage expressions for the mass flux. Even with any of the derived interfacial heat flux expression, I will show that there are still not sufficient equations to solve for all the unknowns when combined with continuum descriptions on the two phases.

Experimentally, Ward and co-workers [7,8] measured the temperature profile of a liquid-vapor interface and noticed that quite a large temperature discontinuity exists at the interface. Although most solutions of the BTE will predict a temperature discontinuity, no theory seems to have been able to predict the magnitude of the measured temperature discontinuity [9,34]. Ward and co-workers developed a statistical rate theory and the corresponding evaporation rate to replace the Schrage equation, but the theory still has the same difficulties in terms of not having enough equations to solve for the unknowns [35]. Bond and Struchtrup [34] used kinetic theory with an interfacial temperature discontinuity to model Ward's experiment, but their model showed much smaller temperature discontinuity than the experiments, although the agreement with experiments became better after including surface curvature. Follow-up experiments by several other groups also showed the interfacial temperature discontinuities [9–14], as well summarized in Ref. [23] . Badam et al. [9] used the temperature discontinuity boundary condition of Cipolla et al. [20], but found that the modeling results were 10-20 times smaller than their experimental data.

The lack of favorable comparison between experiments and modeling, coupled to difficulties in the experiments and modeling, raises doubt on the existence of the interfacial temperature jumps. However, discontinuities at interfaces are rules rather than exceptions in different fields. Well-known examples are velocity and temperature slips in rarefied gas flow [36], phonon interfacial thermal resistance [37], the Deissler temperature jump boundary condition for thermal radiation [38], and interfacial voltage drops. These discontinuity conditions provide a convenient way to connect transport through the interfacial region with the bulk region. Different ways to derive the discontinuity conditions had been developed, as the starting points of the transport equations are usually different. However, for most dilute particles, kinetic theory based on the BTE provides a common starting point [39,40]. I had shown before that one can consistently derive the interfacial discontinuity conditions for different carriers starting from the diffusion approximation to the BTE, which I called diffusion-transmission boundary conditions, and applied the strategy to study rarefied fluid flow, heat conduction, and electrical contact resistance [41,42].

In this work, I will follow same strategies to derive a set of interfacial conditions that are sufficient to connect the Knudsen layer with the liquid and vapor continuum phases, which include, in addition to the well-known Schrage equation, a relation between the interfacial heat flux and temperature discontinuity, as well as an expression for the interfacial density discontinuity (Sec. 3). With these interfacial relations, one can connect continuum descriptions for the liquid and the vapor-phases and solve for the temperature, density, and pressure of both phases immediately near the interface. These solutions will give interfacial temperature, density, and pressure jumps. I will show the usefulness of these interfacial conditions by first demonstrating that these conditions can well explain the past experimental data on the interfacial temperature discontinuities [7,9]. (Sec. 4.1). Since most evaporation and condensation heat transfer experiments report an overall evaporating heat transfer coefficient, I will connect the overall heat transfer coefficient to the single-phase convective heat transfer coefficient. Such modeling suggests that the interfacial temperature discontinuities are not negligible for practical heat transfer scenarios (Sec. 4.2). Before deriving interfacial discontinuity conditions, I will provide a thermomolecular emission model for the transport between the liquid and vapor phases (Sec. 2), mimicking the well-established thermionic emission theory for electron transport across interfaces. Although the final interfacial conditions do not depend on this model, I believe the thermomolecular emission model provides useful insights on the interfacial phase-change process and the origin of the evaporation and condensation coefficients that are often used in literature. Because this paper includes many derivations, I will put most of these derivations in supplemental materials (SM) for interested readers, while focusing the main text on key results.

## 2. THERMOMOLECULAR EMISSION MODEL FOR EVAPORATION

I use the phrase "thermomolecular emission" to mimic the "thermionic emission" that was discovered by Edmond Becquerel in 1853 [43] and rediscovered by Thomas Edison in 1880 [44]. Thermionic emission refers to the evaporation of electrons from hot metal surface as shown in Fig.1(a). Most mobile electrons have energy near the Fermi level. Those electrons with kinetic energy larger than the workfunction can escape the metal surface, entering the surrounding vacuum. The correct theory for thermionic emission was developed by Owen Richardson [45,46]. The development of the thermionic emission theory invoked the physical picture behind liquid thermal evaporation. Ironically, I have not found a similar theory for the liquid thermal evaporation. In this section, I will present such a theory, and call it the thermomolecular emission theory in analogy to the thermionic emission. The thermionic emission model is also often the basis for extension to include other processes such as photon-enhanced thermionic emission [47]. The thermomolecular emission model presented below might also serve as a starting point to model the photomolecular evaporation processes that we recently discovered [48,49].

Thermal evaporation bears a similar picture to thermionic emission, as shown in Fig. 1(b). Molecules in the liquid are bound together near its average potential energy, which is lower than the vapor phase (analogous to vacuum) by $\Delta$, the latent heat per molecule. The water latent heat at 25 $^{o}$C is 2346.3 kJ/kg, which is equivalent to 0.44 eV/molecule. The kinetic energy of molecules spread from zero to infinite. These molecules in the interfacial region with kinetic energy larger than $\Delta$ can escape liquid and enter the vapor phase, as suggested also by molecular dynamics simulations [50].

A key challenge in modeling liquid is to specify the statistical distribution function. Unlike dilute gases which can be described by the one-particle Maxwell-Boltzmann distributions, liquid molecules are strongly correlated with each other via potential interactions. Liquid molecules also differ from that in crystalline solids, which are periodically arranged [51]. The randomness plus the strong correlation have hindered progress in describing thermal transport in liquid. The BBGKY hierarchy is an approach that progressively reduces the N-particle distribution function (where N is the total number of molecules or atoms) to eventually the one-particle distribution function, i.e., the Maxwell-Boltzmann distribution [52,53],

$$f_L(\boldsymbol{v}) = n_L \left[\frac{m}{2\pi k_B T}\right]^{3/2} exp\left\{-\frac{m[v_x^2+v_y^2+v_z^2]/2}{k_B T(z)}\right\}, \tag{3}$$

where $k_B$ is the Boltzmann constant, m is the molecular mass, T the absolute temperature, **v** the velocity with components $v_x$, $v_y$, $v_z$, and $n_L$ the density of the liquid. Unlike dilute gas, for which the N-particular distribution function is the product of the one-particle distributions, the liquid N-particle distribution functions cannot be constructed from the one-particle distribution function alone. However, for our purpose of considering the evaporation of individual molecules, assuming validity of Eq. (3) for each molecule is a reasonable starting point, since the kinetic energy expression remains the same even for correlated liquid molecules [53]. Using the above distribution function, and assuming only molecules with kinetic energy component in the direction perpendicular to the interface (the z-direction) larger than the energy barrier $\Delta$ can escape the interface, i.e.,

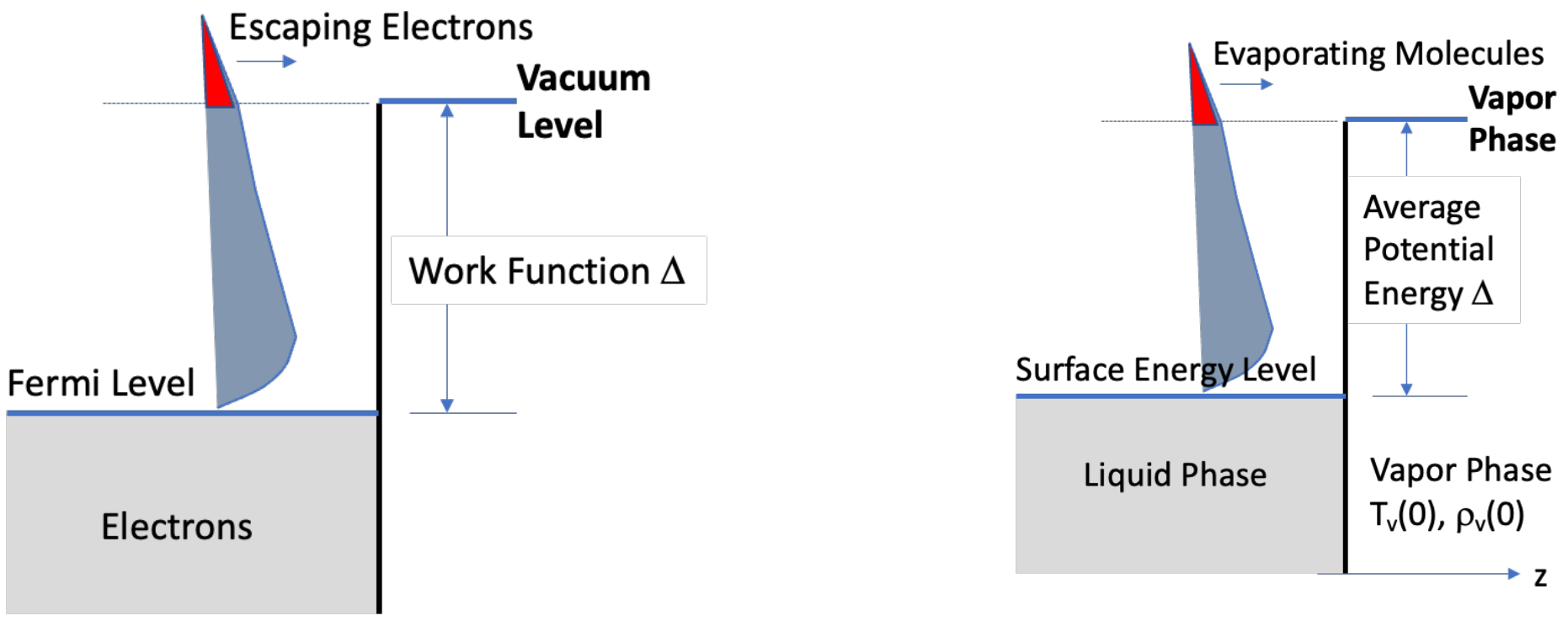

(a) (b)

Figure 1: Illustrations demonstrating similarities between (a) thermionic emission of electrons from a hot surface and (b) thermomolecular emission from a liquid-vapor interface.

$$\frac{1}{2}mv_z^2 \geq \Delta, \tag{4}$$

the flux of the molecules that can leave surface is (see SM, Sec.1)

$$J_m^+ = \varepsilon n_L \sqrt{\frac{k_B T_l}{2\pi m}} exp\left\{-\frac{\Delta}{k_B T_l}\right\}. \tag{5}$$

In the above expression, $\varepsilon$ is artificially added to include nonidealities, representing the molecular emittance from the liquid phase. Note that $\varepsilon$ is not the evaporation coefficient as used in the evaporation literature [33] because here $n_L$ is the liquid density while the evaporation coefficient is defined based on the vapor density. Its relationship with the evaporation coefficient will be discussed later. Detailed math behind arriving at Eq. (5) is given in the SM, Sec. 1, in which I also give an expression when the condition for molecules leaving the surface is set as the total kinetic energy larger than the energy barrier. Richardson's original work did consider this latter case, although the classical expression for thermionic emission used in electronics mostly imposes the condition of normal component of kinetic energy larger than the energy barrier as Eq. (4).

The above expression describes molecules leaving the liquid-vapor interface. Meanwhile, molecules from the vapor phase also moves towards the interface. The Maxwell-Boltzmann distribution, Eq. (3) can be more easily justified for molecules in the vapor phase (it does imply the ideal gas approximation). We consider vapor molecules adjacent to the liquid surface and neglect the collisions among the molecules themselves. Some of these molecules move towards to liquid surface. The vapor molecular flux absorbed by the liquid surface can be similarly obtained as (SM, Sec. 1)

$$J_m^- = \alpha n_v \sqrt{\frac{k_B T_v}{2\pi m}}, \tag{6}$$

where $n_V$ is the density of the vapor-phase molecules and $T_v$ its temperature, $\alpha$ represents the fraction of incoming vapor flux absorbed by the liquid phase, i.e., the condensation coefficient.

In deriving Eq. (6), we assumed that the vapor molecules near the interface obey the Maxwell-Boltzmann distribution, and yet they reach the interface without any collision. These two assumptions conflict each other intrinsically in the following sense. The Maxwell-Boltzmann distribution means molecules are in thermal equilibrium, which can only be reached if there are collisions among molecules. Hence, Eq. (6) is at best an approximation where $T_v$ is considered as the vapor temperature at the edge of the Knudsen layer, but any collision within the Knudsen layer is neglected. Extensively studies had been reported in the past treating collisions within the Knudsen layer via solving the Boltzmann transport equation, Monte Carlo simulations, and molecular dynamics simulations, as cited before. In these studies, $T_v$ is usually considered as a boundary condition at the outer edge of the Knudsen layer. However, we should bear in mind that even at the edge of the Knudsen layer where continuum description is considered valid, the transport is still at nonequilibrium.

At equilibrium, i.e., when $T_l$=$T_v$, the outgoing and incoming flux are balanced $J_s^+ = J_s^-$, which leads to

$$\frac{\alpha}{\varepsilon} = \frac{n_l}{n_s} exp\left\{-\frac{\Delta}{k_B T_l}\right\}, \tag{7}$$

where we use $n_s$ to represent $n_v$ at equilibrium with the liquid, i.e., the saturation condition. This is an example of the detailed balance principle, which in fact applies not only to the total flux integrated over all directions as we used here, but also along each direction. Because of the detailed balance principle, the outgoing flux can also be expressed using the vapor phase properties:

$$J_m^+ = \alpha(T_l) n_s(T_l) \sqrt{\frac{k_B T_l}{2\pi m}}. \tag{8}$$

In the above expression, I was careful to denote that $n_s$ and $\alpha$ should be taken at the temperature of the liquid surface. In literature, $\alpha$ in Eq. (8) is often defined as the evaporation coefficient [33]. The detailed balance principle, i.e., at equilibrium, there is no net flux, guarantees that the evaporation coefficient equals the condensation coefficient or the accommodation coefficient. When the evaporation coefficient is directionally and energetically dependent, the detailed balance principle guarantees that the evaporation and condensation coefficients are equal for every direction and molecular velocity.

When $T_l \neq T_v$, the net molecular mass flux is the difference between $J_m^+$ and $J_m^-$,

$$J_m = \alpha(T_l) n_s(T_l) \sqrt{\frac{k_B T_l}{2\pi m}} - \alpha(T_v) n_v(T_v) \sqrt{\frac{k_B T_v}{2\pi m}} \approx \alpha \sqrt{\frac{1}{2\pi M R}} \left( \frac{P_s(T_l)}{\sqrt{T_l}} - \frac{P_v(T_v)}{\sqrt{T_v}} \right), \quad (9)$$

where in the last step we have replaced the molecular density with the ideal gas relation $P = k_B n T$. The above relation is Hertz-Knudsen equation given in Eq. (1). In arriving at this equation, we assumed that the condensation coefficient is independent of temperature. [NO_PRINTED_FORM]In reality, the condensation coefficient depends strongly on temperature, as we will show below.

The use of detailed balance principle enables us to express the emission of molecules from liquid in terms of vapor properties as in Eq. (8). This approach was also taken by Herring and Nichols in their treatment of thermionic emission [54]. It is also well-established in thermal radiation where we use the blackbody radiation law in vacuum rather than inside the emitting medium, with $\alpha$ similar to emissivity, and its equality to the condensation coefficient similar to the Kirchhoff's law [55]. However, the treatment of thermomolecular emission process from liquid phase sheds light on the physical pictures of thermal evaporation, which we will discuss below.

First, the Gibbs phase rule says that for pure substance at liquid-vapor phase equilibrium, only one intensive property, for example temperature, is changeable. For a given temperature, the saturation pressure and the liquid and the vapor phase molecular densities are all fixed. We use saturated water data and set $\Delta = ML/N_A$, where $N_A$ is the Avogadro constant, L the latent heat, and obtain the ratio of $\alpha/\varepsilon$ according to Eq. (7) based on water properties. This ratio is shown in Fig. 2(a). One can see that $\alpha/\varepsilon$ depends strongly on temperature. This strong temperature dependence hints that α itself is strongly temperature dependent and explains why often one needs to separate the evaporation coefficient from the condensation coefficient.

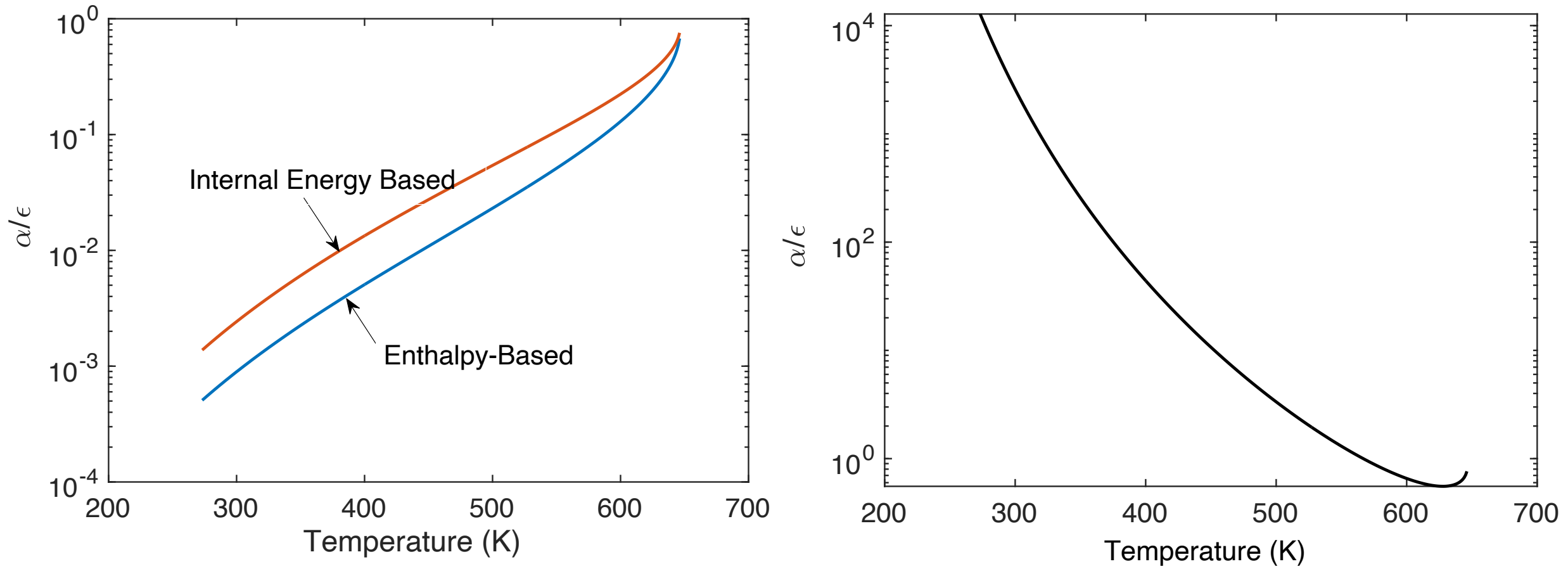

Figure 2. Ratio of the vapor accommodation coefficient and the emittance [defined according to Eq. (5)] based on (a) internal energy and enthalpy of phase change and (b) after considering surface energy assuming interfacial density changes occur only over one molecular layer, using properties of water.

Second, in the above calculation, we assumed that the latent heat is the energy barrier needed to overcome. This picture is not entirely correct, however, since molecules at the surface have higher energy than that of the molecules inside, while the latent heat per molecule is averaged to all molecules in the domain and dominated by molecules in the bulk region. Suppose that the energy per molecule at the surface is σ, then the energy barrier needed to overcome for evaporation is Δ-σ. Evaluating σ, however, is difficult, as the density of interfacial molecules change from bulk liquid to vapor phase over a distance of 3-7 Å [56,57], i.e., involving several molecular layers, and the measured surface tension includes contributions from this region. I attempted to attribute the measured water surface tension to one molecular layer of water on the surface, and replace Δ in Eq. (7) by Δ-σ, the obtained $\alpha/\varepsilon$ ratio is shown in Fig. 2(b). In this case, α becomes much larger than ε and the temperature dependence becomes different, which is probably because I attributed the surface tension to only one molecular layer. In reality, several molecular layers of water molecules contribute to the surface tension.

Also unknown is the distance $\delta$ within liquid itself for the liquid molecule energy to change from bulk value $\Delta$ to their values at surface $\Delta$-$\sigma$. Over this distance, the liquid is cooled, but this cooling is coupled to heat conduction within the liquid phase and the vapor phase. There seems to be no study on this distance. However, surface tension theory suggests that molecules in the surface region experience anisotropic tension (or pressure) [58]. Within the same $\delta$, the pressure changes from isotropic for molecules in the bulk region to anisotropy pressure in the surface region.

When molecules leave the liquid-vapor interface, they also experience an attractive force from the bulk dense molecules in the surface region and inside the bulk region. This force is sometimes attributed to the van der Waals force but is more generally called Casimir force, characterized by the Hamaker constant [59]. The van der Waals force is an electrostatic picture while the Casimir force is based on the electrodynamic picture, typically modelled using the fluctuating electrodynamic approach. This is the origin of the disjoining pressure [59]. A valid question is if the disjoining pressure should be included in phase change modeling.

For evaporation from a bulk liquid surface, the equilibrium condition between the bulk liquid and its vapor phase already included the disjoining pressure caused by the above-mentioned force. This force will affect values of the evaporation and condensation coefficients. However, for a very thin liquid film on a solid substrate, molecules in the vapor phase near the liquid surface will experience a different force due to the long-range interaction with the solid substrate materials. In this case, correction to this different interaction should be made, under the concept of disjoining pressure [59]. Rigorous calculation of this disjoining pressure can be done based on the fluctuational electrodynamics used for calculating the Casimir force. This also explains why accommodation coefficient depends on materials.

Another question is if the latent heat should be used, or the internal energy difference between the vapor and the liquid phase should be used in estimating $\Delta$. The latent heat is the enthalpy change between the liquid and the vapor phase, which includes the work done by volume expansion. Although the above discussion was based the latent heat, i.e., enthalpy, in estimating $\Delta$, using internal energy difference during phase transition to estimate $\Delta$ would better reflect that it represents the potential energy difference. Fig. 1(b) includes $\alpha/\varepsilon$ ratio based both the enthalpy and the internal energy changes during the liquid-vapor phase transition.

Clearly, the above discussions are qualitative. Based on the water data (energy, enthalpy, and surface tension), it seems that both $\varepsilon$ and $\alpha$ depend strongly on temperature. Although past literature had emphasized evaporation coefficient could be different from condensation coefficient, the above discussion suggests what we really need is the temperature dependence of the condensation coefficient. This view was also indicated in the discussion of Ref. [60].

## 3. Diffusion Approximation with Diffusion-Transmission Jump Boundary Conditions

Since even the emission process can be described by the vapor phase properties using the detailed balance principle, in this section we will focus on one-dimensional transport in the vapor phase along the direction perpendicular to the interface, i.e., the z-direction. This problem has been treated in many past studies based on the kinetic theory [15–18,21,61] . Our focus is on deriving interfacial jump boundary conditions. We use the BTE as the starting point. For the bulk region, we will use the diffusion approximation derived from BTE. For the interfacial region, we derive temperature, pressure, and density jump boundary conditions that are consistent with the diffusion approximation in the bulk region. The starting point is the BGK approximation of the BTE [39,40]

$$v_z \frac{df}{dz} = -\frac{f-f_d}{\tau} \tag{10}$$

where $\tau$ is the relaxation time, f the distribution function in the phase space, $f_d$ is the displaced Maxwell velocity distribution[40,62],

$$f_d(z, \boldsymbol{v}, u(z)) = n(z)\left[\frac{m}{2\pi k_B T(z)}\right]^{3/2} exp\left\{-\frac{m\left[v_x^2+v_y^2+(v_z-u(z))^2\right]}{2k_B T(z)}\right\} \tag{11}$$

where u(z) is the average velocity in the z-direction, and n(z) and T(z) are the local density and temperature, respectively. The above expression considers kinetic energy of the molecules only. Extension to include other forms of energy (rotational and vibrational) can follow similar treatment in literature [16], which in the continuum regime will be mainly reflected in using the appropriate specific heat. Although the Maxwell-Boltzmann distribution implies the ideal gas law $P(z) = k_B T(z) n(z)$, we will delay this substitution. This way, the expression may be more general and applicable to cases where the ideal law is not obeyed.

### 3.1. Diffusion Approximation for the Bulk Region

Away from the interface, the first order perturbation solution to the BTE leads to the distribution function as

$$f = f_d - \tau v_z \frac{df_d}{dz} = f_d - f_d \tau v_z\left\{\frac{1}{n}\frac{dn}{dz} + \frac{1}{T}\frac{dT}{dz}\left[\frac{m\left[v_x^2+v_y^2+(v_z-u(z))^2\right]}{2k_B T(z)} - \frac{3}{2}\right] + \frac{m[v_z-u(z)]}{k_B T(z)}\frac{du}{dz}\right\} \tag{12}$$

We consider that the above expression is valid outside the Knudsen layer. The density, velocity, and temperature are all local equilibrium values (and correspondingly the pressure). We will use Eq. (12) to derive transport equations for the continuum region, away from the interface.

From Eq. (12), we can derive expressions for the mass flux $\dot{m}$ and heat flux q as (see SM, Sec. 2 for details):

$$\dot{m} \approx \rho u - \frac{RT}{M}\tau\frac{d\rho}{dz} - \frac{RT}{M}\rho\tau\frac{1}{T}\frac{dT}{dz} = \rho u - L_{11}\frac{d\rho}{dz} - L_{12}\frac{dT}{dz} \tag{13}$$

$$q \approx \frac{5R}{2M}T\rho u - \frac{5}{8}\tau\left(\frac{2RT}{M}\right)^2\frac{d\rho}{dz} - \frac{5}{4}\rho\tau\left(\frac{2RT}{M}\right)^2\frac{1}{T}\frac{dT}{dz} = \frac{5R}{2M}T\rho u - L_{21}\frac{d\rho}{dz} - L_{22}\frac{dT}{dz} \tag{14}$$

We use the approximate signs to indicate that higher order nonlinear terms are neglected. These terms are given in the SM Sec.2 and can be included if the Mach number is high. The three terms in the above expressions are due to (1) convection, (2) self-diffusion from the density gradient, and (3) diffusion due to the temperature gradient. The density and the temperature gradients both contribute to the mass and heat transfer, and the cross terms represented by $L_{12}$ and $L_{21}$ are related, a manifestation of the Onsager reciprocity relation.

We can write Eq. (13) as

$$\frac{d\rho}{dz} = -\frac{\dot{m} - \rho(z)u(z)}{\frac{RT}{M}\tau} - \frac{\rho}{T}\frac{dT}{dz} \tag{15}$$

Note that the continuity condition for one-dimensional transport leads to,

$$\rho(z)u(z) = constant \tag{16}$$

Substituting Eq. (15) into Eq. (14), we have

$$q = \frac{5}{2}\frac{RT}{M}\dot{m} - \frac{5}{8}\rho\tau\left(\frac{2RT}{M}\right)^2\frac{1}{T}\frac{dT}{dz} = \frac{5}{2}\frac{RT}{M}\dot{m} - k\frac{dT}{dz} \tag{17}$$

where k is the thermal conductivity. Equations (13-17) are applicable to the bulk region. We need connect them to the interfacial region, which we will treat in the next section.

In literature, the mass flux $\dot{m}$ is often set to equal to $\rho u$, i.e., the first term of Eq. (13). Or alternatively, one can argue that the velocity itself is defined based on Eq. (13), including the temperature and density gradient terms. The former treatment completely neglects the diffusion process under temperature and density gradients. In the case of pure heat conduction, u=0 and one can use the mass flux equaling zero to eliminate the density gradient in arriving at the proper thermal conductivity as given in Eq. (17) [39,40]. However, when there is macroscopic convection, it is unlikely that the two diffusion terms in Eq. (13) cancel each other. How to treat such self-diffusion is a question still under debate [63–68]. Brenner introduced a mass velocity and a volume velocity and modifications to the Navier-Stokes equations [63,64]. Similar treatment was introduced by Elizarova [[67]. The treatment presented here is consistent with their views. Equation (16) is the consequence of ensuring that density n(z) represents the local density in the continuum (see SI, Sec.2).

### 3.2 Interfacial Mass Flux

Consider a stationary interface and transport in the interfacial region, applying detailed balance principle to each direction, the distribution of the molecules leaving the surface can be expressed in terms of a Maxwellian distribution as

$$f_s^{+}(T_l) = \alpha(T_l) n_s(T_l) \left[\frac{m}{2\pi k_B T_l}\right]^{3/2} exp\left\{-\frac{m[v_x^2+v_y^2+v_z^2]}{2k_B T_l}\right\} \quad \text{(for } v_z>0\text{)} \qquad (18)$$

We have assumed an isotropic accommodation coefficient. Note that unlike the vapor phase which obeys the drifted Maxwellian distribution, molecules leaving the liquid phase have a much smaller velocity, which is neglected in Eq. (18).

In the diffusion-transmission approach to derive the interfacial discontinuities [41], the distribution of molecules coming towards the interface is identical to that used for the diffusion approximation, but only for $v_z<0$ and at z=0, i.e., the density, temperature, and velocity in Eq. (12) are the values in the vapor phase at z=0. These approximations are inherently similar to those used in the Hertz-Knudsen or the Schrage equations, implying that we neglect the finite thickness and collisions within the Knudsen layer that eventually lead to the molecular distribution following the diffusion approximation Eq. (12). The values of $T_v(0)$, u(0), and $n_v(0)$ are actually values at the outer edge of the Knudsen layer. Many of the past treatments of the Knudsen layer also assumed that the molecules outside the Knudsen layer obey the displaced Boltzmann distribution, i.e., Eq. (11), with however zero gradients of n, T, and u beyond the Knudsen layer. Such zero gradients are inconsistent with continued diffusion in the continuum region as represented by Eqs. (13)-(17). In contrast, the use of Eq. (12) to represent the distribution of molecules coming towards evaporating interface ensures that interfacial conditions we derived below are consistent with diffusion in the continuum. Hence, we expect that the ballistic transport approximation within the Knudsen layer incurs inaccuracy compared to models rigorously treating the Knudsen layer, but gains accuracy in connecting with the continuum treatment.

Among the incoming molecules, only a fraction $\alpha(T_v(0))$ is absorbed similar to Eq. (6). The rest fraction 1- $\alpha(T_v(0))$ is reflected. Subtle questions are what the distribution of the reflected molecules is and what the equivalent temperature of these reflected molecules is. The most likely process is that molecules get absorbed and then re-emitted. But if that is the case, it means an accommodation coefficient equaling one. In fact, most molecular dynamics simulations suggest indeed that accommodation coefficient is close to one, although experimental values vary several orders of magnitude [60]. To keep an accommodation coefficient in the theory, probably the next good assumption is diffuse scattering. That is, the incoming molecular flux is redistributed isotropically into the hemispherical direction. It is also possible to assume some specular reflection, i.e., the reflected molecules will leave surface in mirror-like fashion. My derivation will use this latter approximation for mathematical simplicity, although physically, diffuse scattering is more likely. For transport in the perpendicular direction, my sense is that the difference between the two approximations should be relatively small, especially when $\alpha$ is close unity. Molecular dynamics simulations do show that kinetic energy distributions can be anisotropic as characterized by anisotropic temperatures [26,27]. Specular reflection approximation would lead to some kinetic energy anisotropy that is consistent with such simulations.

Using $f_s^+(T_l)$ in Eq. (18) for $v_z>0$, we can express the outgoing mass flux from the interface as (SM, Sec. 1)

$$J_m^+ = \alpha n_s(T_l)\sqrt{\frac{k_B T_l}{2\pi m}} \tag{19}$$

and using Eq. (12), the molecular flux coming towards the interface is (SM, Sec.3, Eq.(A33))

$$J_m^- = \alpha\left[-n\left(\frac{k_B T_v(0)}{2\pi m}\right)^{1/2} + \frac{J_m}{2}\right] \tag{20}$$

Note that the above expression contains the unknown net molecular flux $J_m$, since the diffusion creates a distortion of the molecular distribution function as is evident in Eq. (12). The difference between $T_l$ and $T_v(0)$ represents the temperature jump at the interface. The sum of $J_m^+$ and $J_m^-$ (which is negative) gives the net molecular flux

$$J_m = \alpha n_s(T_l)\sqrt{\frac{k_B T_l}{2\pi m}} + \alpha\left[-n\left(\frac{k_B T_v(0)}{2\pi m}\right)^{1/2} + \frac{J_m}{2}\right]$$

Solving for $J_m$, we obtain expression for the molecular flux

$$J_m = \frac{2\alpha}{2-\alpha}\left[n_s(T_l)\sqrt{\frac{k_B T_l}{2\pi m}} - n(T_v)\left(\frac{k_B T_v}{2\pi m}\right)^{1/2}\right]$$

Or in terms of the mass flux

$$\dot{m} = \frac{2\alpha}{2-\alpha}\sqrt{\frac{R}{2\pi M}}\left[\rho_s(T_l)\sqrt{T_l} - \rho_v(0)\sqrt{T_v(0)}\right] \tag{21}$$

$$= \frac{2\alpha}{2-\alpha}\sqrt{\frac{M}{2\pi R}}\left[\frac{P_s(T_l)}{\sqrt{T_l}} - \frac{P_v(0)}{\sqrt{T_v(0)}}\right] \tag{22}$$

The last step used the ideal gas law. Although the ideal gas law is implied in the Maxwell-Boltzmann distribution, in practice, using Eq. (21) is more accurate for liquids such as water. The last step is identical to the Schrage expression, although he arrived at the same expression using a different method [32,69]. It is interesting that Schrage's derivation included only the drift velocity but did not include the temperature nor the density gradients, while what we presented here included all three terms, and yet the final results are the same because contributions from both the density and temperature gradients to the distortion of the distribution function weigh equally as the drift contribution (ρu) he considered.

### 3.3. Interfacial Heat Flux and Temperature Discontinuity

Equations (21) and (22) suggest that at the liquid-vapor interface, discontinuities in the temperature, the density, and the pressure are possible. Schrage's equation only gave the mass flux, which is not sufficient to determine the discontinuities. Here, we use the same strategy, i.e., using Eq. (12) to represent the molecules coming towards the interface, and derive an interfacial heat flux expression.

The heat flux coming towards the interface is [SM, Sec. 3, Eq. (A44)]

$$q^{-} \approx -\frac{m}{2\sqrt{\pi}} n \left(\frac{2k_B T}{m}\right)^{3/2} + \frac{q}{2} \tag{23}$$

Again, $q^{-}$ itself contains the net heat flux q, which equals the sum of $q^{+}$ and $q^{-}$. The outgoing heat flux due to surface emission can be expressed as [SM, Sec. 3, Eq.(A45)]

$$q^{+}(z) = \alpha \int_{-\infty}^{\infty} dv_x \int_{-\infty}^{\infty} dv_y \int_{0}^{\infty} v_z \frac{mv^2}{2} f_e dv_z + (1-\alpha) q^{-}$$

$$= \frac{m}{2\sqrt{\pi}} \alpha n_s(T_l) \left(\frac{2k_B T}{m}\right)^{3/2} - (1-\alpha) q^{-} \tag{24}$$

where the first term is due to emission and the second term represents the reflection of the incoming flux. Summing up Eqs. (23) and (24), we get

$$q = \frac{m}{2\sqrt{\pi}} \alpha n_s(T_l) \left(\frac{2k_B T_l}{m}\right)^{3/2} + \alpha \left(-\frac{m}{2\sqrt{\pi}} n(T_v) \left(\frac{2k_B T_v}{m}\right)^{3/2} + \frac{q}{2}\right)$$

Rearranging the above equation, we get

$$q = \frac{2\alpha}{2-\alpha} \frac{R}{M} \sqrt{\frac{2R}{\pi M}} \left[\rho_s(T_l) {T_l}^{3/2} - \rho_v(0) {T_v}^{3/2}\right] \tag{25}$$

$$= \frac{2\alpha}{2-\alpha} \sqrt{\frac{2R}{\pi M}} \left[P_s(T_l) {T_l}^{1/2} - P_v(0) {T_v}^{1/2}\right] \tag{26}$$

where $T_v=T_v(0)$ is again the temperature of the vapor phase immediately at the interface. We can use Eqs. (21) and (22) to further express the heat flux as

$$q = \frac{2RT_l}{M} \dot{m} + \frac{2\alpha}{2-\alpha} \frac{R}{M} \sqrt{\frac{2R}{\pi M}} \rho_v(0) \sqrt{T_v} (T_l - T_v) \tag{27}$$

$$= \frac{2RT_l}{M} \dot{m} + \frac{2\alpha}{2-\alpha} \sqrt{\frac{2R}{\pi M}} \frac{P_v(0)}{\sqrt{T_v}} (T_l - T_v) \tag{28}$$

Equations (25)-(28) are equivalent. Sone and Onishi derived a temperature jump boundary condition but they had a different form [18]. Our form is similar to that of Bond and Struchtrup [34], although their coefficients are more complicated since they employed a model of condensation coefficient that depends on energy. In their work, diffuse and specular reflection is considered, where the diffuse reflection is assumed to obey the Maxwell distribution with a different equivalent pressure and liquid temperature. At equilibrium, they further proved that the pressure for diffuse evaporation must be the saturation pressure, using the detailed balance principle. However, during nonequilibrium, they calculated a different equivalent pressure for the diffusely reflected molecules. Their Eq. (44) is identical to what I presented above.

Barrett and Clement [33] actually derived a similar equation as Eq. (25), again without considering the diffusion terms. However, they argued that this equation is in conflict what one will get by multiplying the mass flux as given by the Schrage equation with the enthalpy of the vapor 5RT/2. Due to this conflict, they stated that the Schrage equation violates energy and momentum conservation laws, hence casting doubts on the Schrage equation. I cannot agree with Barrett and Clement for the following reasons. From the above derivations for both the Schrage equation and the interfacial heat flux expression, it is clear that the interfacial region is in highly nonequilibrium states. One cannot easily define a temperature. An enthalpy of $5RT_v/2$ represents the local equilibrium temperature, while the molecules going in the positive and the negative z-directions are clearly at different temperatures and carry different enthalpies. Mathematically, one cannot write the product of two quantities in the phase-space integral, see for example Eq.(A34) which is the product of molecular flux and molecular kinetic energy, as the product of two averaged quantities.

Another fine detail is also worth mentioning. The above discussion clear defines that the interfacial temperature jump as represented by Eq. (27) is between the liquid surface temperature and that of the vapor temperature at the outer edge of the Knudsen layer, despite that we approximated this outer edge as z=0. In kinetic approaches solving the Boltzmann transport equations or molecular dynamics simulations [27,70–72], this temperature jump is replaced by a continuous varying temperature distribution over the Knudsen layer and hence there may be no apparent temperature jump at physical liquid-vapor interface. However, sometimes one can still discern a small temperature discontinuity at the liquid-vapor interface. This is because the angular distribution of the evaporated molecules is usually assumed to be Maxwellian, i.e., isotropic, which is qualitatively different from the distributions of the molecules near the interface. This asymmetric can lead to the equivalent temperature of the vapor phase different from that of the liquid. The interfacial temperature drop in our heat flux equation represented by Eq. (27) includes this temperature drop as well as the temperature drop throughout the Knudsen layer.

### 3.4 Interfacial Density Jump

While it is recognized that there is interfacial density discontinuity at evaporation, as is seen from solutions of the Boltzmann transport equation, I have not found reports on interfacial density jump boundary condition. Here, I will use the same strategy to derive an interfacial density discontinuity boundary condition.

The molecular number density $n_v(0)$ can be determined by

$$n_v = \int_{-\infty}^{\infty} dv_x \int_{-\infty}^{\infty} \left[\int_{-\infty}^{0} f^- dv_z + \int_{0}^{\infty} f^+ dv_z\right] \quad (29)$$

where f⁻ is gain given by Eq. (12) and f⁺ consists of $f_s^+(T_l) + (1-\alpha)f^-$. Substituting these expressions and carrying out the integration (SM, Sec. 3), we have

$$n_v(0) = \frac{n_v(0)}{2} + \frac{1}{2}\tau\frac{dn}{dz}\left(\frac{2k_BT}{\pi\, m}\right)^{1/2} + \frac{3}{4}n\tau\frac{1}{T}\frac{dT}{dz}\left(\frac{2k_BT}{\pi m}\right)^{1/2}$$

$$+\alpha\frac{n_s}{2} + (1-\alpha)\left[\frac{n_v(0)}{2} + \frac{1}{2}\tau\frac{dn}{dz}\left(\frac{2k_BT}{\pi\, m}\right)^{1/2} + \frac{3}{4}n\tau\frac{1}{T}\frac{dT}{dz}\left(\frac{2k_BT}{\pi m}\right)^{1/2}\right] \quad (30)$$

which can be written as

$$n_v(0) = n_s + \frac{(2-\alpha)}{\alpha}\tau\left(\frac{2k_BT}{\pi\, m}\right)^{1/2}\left\{\frac{dn}{dz} + \frac{3}{2}n\frac{1}{T}\frac{dT}{dz}\right\} \quad (31)$$

We can use Eqs. (13) and (14) to express the above equation as

$$\rho_v(0) = \rho_s + \frac{(2-\alpha)}{\alpha}\left(\frac{2M}{\pi\, RT_v}\right)^{1/2}\left[\rho u - \frac{\dot{m}}{2} - \frac{Mq}{5RT_v}\right] \quad (32)$$

The above expression is explicitly written for the case of evaporation. For the case of condensation, the equation also holds although $\dot{m}$ as given by Eqs. (21) and (22) are negative, while q can be either positive or negative depending on the solution of heat transfer in the vapor phase.

Equation (21) or (22) for the mass flux, one of the Eqs. (25)-(28) for the heat flux, and (31) or (32) for the density discontinuities give three independent equations that can connect the interfacial density at saturation condition determined by $T_l$, the vapor phase temperature $T_v$, density $\rho_v$, or pressure $P_v$ immediately at the interface, i.e., at the outer edge of the Knudsen layer, with the interfacial mass and heat fluxes. These interfacial conditions can be used in conjunction with the continuum descriptions for the vapor and the liquid regions. Eqs. (13) and (14) are constitutive equations for the one-dimensional transport in the vapor bulk phase, neglecting higher order nonlinear effect (which can be included by referring to results in the SM). In the following sections, we will treat the problem of evaporation and condensation at a single interface, which is commonly seen in heat transfer applications.

## 4. EVAPORATION OR CONDENSATION AT A SINGLE INTERFACE

Evaporation or condensation on a single flat interface forms the foundation to many relevant phase-change problems. Ytrehus [21] assumed that there are three groups of molecules: emitted that is determined by the surface temperature, and another forward going flux depending on temperature outside the Knudsen layer multiplied by a coefficient due to reflection of incoming molecules, and a backward flux. He used the moments of the BTE to arrive at equations determining the unknown coefficients. His treatment did not consider the influence of the transport in the bulk region, which is assumed to be of zero gradient. Pao [15] studied the half space problem via solving the integral form of the BTE, assuming a linear temperature distribution far away from the interface. He showed that the problem is identical to the case of rarefied gas conduction when flow is zero: reducing to two integral equations.

In certain sense, the way we established the interfacial conditions contains the Ytrehus picture of three different streams of molecules: emitted, incoming from the bulk region, and reflected, although we assumed ballistic transport in the Knudsen layer, and the scattering effect is only implied from the molecular distribution given in Eq. (12). Since Eq. (12) is also the foundation for transport in the bulk region, the interfacial conditions we established enable us to connect to different transport conditions in the continuum region. Here, we will first consider evaporation and condensation above a liquid film so that we can compare with previous experiments. We will then connect typical heat transfer measurement, which usually reports an overall evaporation heat transfer coefficient, to the interfacial discontinuities. We assume that the water layer thickness is fixed, i.e., there is water supply to the layer in case of evaporation and extraction from the layer in case of condensation. We assume the water layer is thick enough so we can neglect the disjoining pressure. We also assume that the absorption of the latent heat happens in the liquid phase and right at the surface, although as discussed in Sec. 2, heat is absorbed over a certain distance close to the interfacial region when molecules move from inside the bulk liquid to the surface region.

### 4.1 Temperature Discontinuity at An Evaporating or Condensing Interface: Comparison with Experiment

Ward and co-workers carried out pioneering experiments demonstrating temperature discontinuities at a water-vapor interface [7,8]. In their experiment, the evaporation rate is controlled externally by a vacuum pump. They measured that the liquid side temperature is lower than the vapor side. Badam et al. [9] conducted similar measurements, with additional data showing even larger temperature discontinuity when the vapor phase is heated. They listed more details of their experiments, including the liquid and vapor side heat fluxes, although the vapor side heat flux is calculated from a model. I will focus on comparing with Badam et al.'s experiment because of their more detailed information and the fact that their interface is flat, while the liquid-vapor interface is curved in Ward and co-workers's experiments.

As shown in Fig. 3(a), we can divide the problem into three regions: the bulk vapor region (region I), the interfacial region (region II), and the liquid film (region III). We first consider the bulk region, for which Eqs. (13)-(17) are applicable. In a typical experimental configuration, the pressure above the surface is kept at certain value $P_{v,\infty}$ using a vacuum pump. So, the boundary condition can be written as

$$z = d_v, \;\; P_v = P_{v,\infty} \tag{33}$$

We first consider transport in the bulk vapor phase. From Eq. (17) and considering that the relaxation time expression dependence on density

$$\tau = \frac{m}{\pi\sqrt{2}\rho a^2 v_{th}} \tag{34}$$

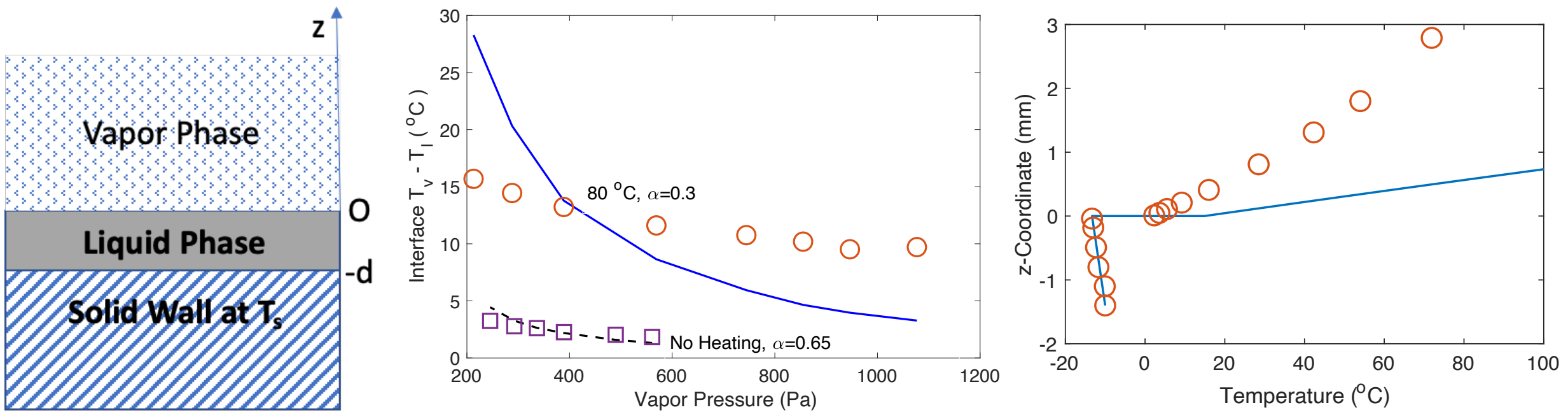

(a) (b) (c)

Figure 3. (a) Illustration of evaporation or condensation at a single interface. Comparison of (b) calculated (lines) and measured (dots) interfacial temperature jump and (c) temperature distributions (experimental data from Badam et al. [9]).

we see that thermal conductivity depends $T^{1/2}$. In Eq. (34), a is the diameter of the molecule. To include this temperature dependence, we can rewrite Eq. (17) as

$$0 = \frac{5}{2}\frac{R\dot{m}}{M}\left(T - \frac{2Mq}{5R\dot{m}}\right) - k_o\left(\frac{T}{T_o}\right)^{1/2}\frac{dT}{dz} \tag{35}$$

The above equation can be analytically solved for the temperature distribution in the bulk vapor phase. The solution is given in SM Sec. 4. We can also approximate the thermal conductivity as a constant, and in this case, the temperature distribution is an exponential function.

$$\left[\frac{\frac{2Mq}{5R\dot{m}} - T(z)}{\frac{2Mq}{5R\dot{m}} - T_v(0)}\right] = \exp\left(\frac{5}{2}\frac{R\dot{m}}{Mk}z\right) \tag{36}$$

The density or pressure distribution in the vapor phase can be obtained from solving Eq. (15) (SM, Sec.4)

$$ln\left(\frac{T(z)\rho(z)}{T_v(0)\rho_v(0)}\right) = ln\left(\frac{P(z)}{P_v(0)}\right) = -\frac{(\dot{m}-\rho_\infty u_\infty)}{\rho_\infty D_\infty}\int_o^z \frac{1}{\sqrt{T/T_\infty}}dz \tag{37}$$

Or

$$\frac{P(z)}{P_v(0)} = exp\left\{-\frac{(\dot{m}-\rho_\infty u_\infty)}{\rho_\infty D_\infty}\int_o^z \frac{1}{\sqrt{T/T_\infty}}dz\right\} \tag{38}$$

Note that we do not assume $\dot{m} = \rho_\infty u_\infty$ as discussed before, because Eq. (13) clearly shows that the mass flux depends not only on the bulk motion $\rho_\infty u_\infty$, but also diffusion due to density and temperature gradients.

We now relate the mass flux $\dot{m}$ and heat flux q in the above equations to the interfacial region and the liquid region. In the liquid side, the heat flux that must be taken away is

$$q = k_l\frac{T_w - T_l(0)}{d} - \dot{m}L \tag{39}$$

where $T_w$ is the wall temperature. The first term is heat conduction from the wall to the interface (positive for evaporation and negative for condensation), since the convective heat flux due to liquid flowing in (evaporation) or out (condensation) is small (see SM, Sec. 4 for further justification). The second term is the heat absorbed from liquid-phase due to evaporation, which we assume happens right at the interface. This heat flux should equal to the heat flux carried out by the vapor phase, i.e., the same q used in Eq. (35). The same heat flux goes through interface, which can be represented by one of Eqs. (25-28).

In a typical experiment, $T_s$, $P_{v,\infty}$ are usually known. The unknowns include q, $\dot{m}, u_\infty, T_v(0), \rho_v(0), T_l(0), and\ T_{v,\infty}$ [Eq.(16) always holds due to continuity and ideal gas law will connect ρ, P, and T at the same location]. Since $\rho_s$ corresponds to the saturation density, it can be found either from the steam table or the Clausius-Clapeyron equation once $T_l(0)$ is known. We can use Eqs. (16), (21), (27), (32), (36), (37), and (39), or their alternatives in terms of pressure, to solve for the seven unknows. The problem is well-defined.

Comparison with Badam et al.'s experiments [9] can be done without solving all the above equations simultaneously, since they have listed the liquid and vapor side heat fluxes. As their vapor side heat flux was calculated, we will use liquid side heat flux $q_w$ only. The $q_w$ is essentially the conductive heat flux, i.e., the first term in Eq. (39). Equating Eq. (39) with Eq. (27), we have

$$q_w - \dot{m}L = \frac{2RT_l(0)\dot{m}}{M} + \frac{2\alpha}{2-\alpha}\sqrt{\frac{2R}{\pi M}}\frac{P_v(0)}{\sqrt{T_v(0)}}[T_l(0) - T_v(0)] \quad (40)$$

Badam et al.'s experiments include two scenarios: the vapor phase is heated or not heated. Equation (40) applies for both cases. Figure 3(b) compares the interfacial temperature jump $\Delta T = T_l(0) - T_v(0)$ calculated based on Eq. (40) for both cases, using the measured $\dot{m}$ and $T_l$ in the Badam et al. experiment, and α as a fitting parameter. The interfacial conditions we use here can well explain the large measured interfacial temperature jump. The accommodation coefficient is not far from one.

In Fig. 3(c), we also show the temperature distributions in the liquid and the vapor phase. The temperature distribution in the vapor phase is obtained from Eq. (36). Although my calculation shows similar trend as observed in the experiment, in terms of interfacial temperature jump and temperature distribution trends reversal between the liquid and the vapor phases, the comparison with Badam et al.'s data is not perfect on the vapor temperature distributions. This is understandable since their heater is only 3 mm away from the interface and the heater seems to be made of wires to allow vapor to escape. It is unlikely that transport in the vapor-phase is one-dimensional as we assume here, and the heater region might be much hotter than measured in between heating wires using a thermocouple, as Fig.3(c) suggests.

Although the fact that the measured and simulated temperature distribution seems to be counter-intuitive: the vapor temperature is higher than the liquid interface temperature and heat conducts from both the vapor and the liquid sides to the interface, it is consistent with the fact that evaporation leads to cooling. In the experiments of both Badam et al.[9] and Ward and co-workers [7,8], the liquid layers are thick (~mm range), so the first term in Eq. (39) is not sufficient to supply heat needed for evaporation. Hence, heat conducts from the vapor phase to the liquid-vapor interface. The higher is the evaporation rate, the more heat needs to be supplied from the vapor side, and the larger is the interface temperature jump. This is particularly true in Badam et al.'s experiment when the vapor phase is actually heated. We will show next that in typical heat transfer situations when the liquid layer is thin and wall is heated, the liquid side temperature will be higher than the vapor side. The different trends of the interfacial heat flow were considered a puzzle but are really due to relative magnitudes of liquid heat supply vs. latent heat needed for evaporation, i.e., the two terms in Eq. (39), which determine the sign of q, i.e., if the vapor phase transfers heat to the interface or takes away heat from the interface.

### 4.2 Temperature Discontinuity from Measured Vapor Phase Heat Transfer Coefficient

In evaporation or condensation heat transfer, one often uses a heat transfer coefficient, $h_t$, neglecting details between the liquid and the vapor phases. The heat transfer coefficient is defined based on the wall temperature and bulk vapor phase temperature away from wall ($T_\infty$), or the free stream temperature, i.e.,

$$q_w = h_t(T_w - T_\infty) \tag{41}$$

The question is if we can use the measured evaporative heat transfer coefficient to infer the interfacial temperature discontinuity.

In typical heat transfer situations, the vapor phase transport is usually not one-dimensional as we dealt above. To resolve this difficulty, we use the single-phase convective heat transfer coefficient $h_s$ to describe the convective heat transfer between the vapor phase and the liquid surface instead of computing q from Eq. (17). With this picture, we can write heat transfer in the vapor side as:

$$q = \frac{5}{2}\frac{RT_v(0)}{M}\dot{m} + h_s[T_v(0) - T_\infty] \tag{42}$$

where the first term represents the enthalpy carried by the mass flux and second term due to regular single-phase convective heat transfer. Note unlike the interfacial heat flux, the local enthalpy in the continuum phase is given by the ideal gas specific heat at constant volume $5RT_v/2$. This situation can be treated as coupled heat and mass transport [72].

In addition to using Eq. (42) for the heat flow in the bulk vapor phase, we can similarly replace Eq. (13) for mass diffusion by a mass transfer coefficient $h_m$ so that the mass flux can be expressed as

$$\dot{m} = \rho u + h_m(\rho_v(0) - \rho_\infty) \tag{43}$$

The mass transfer coefficient can be related to the convective heat transfer coefficient by assuming a Lewis number equaling one, with the understanding that this assumption can be easily relaxed for other Lewis number values [72], i.e.,

$$h_m = \frac{h_s}{\rho c_p} \tag{44}$$

Substituting the above expressions for $\dot{m}$ and q into Eq. (32), we get [SM, Sec. 5, Eq.(A70)]

$$\rho_v(0) = \rho_s - \frac{(2-\alpha)}{\alpha}\left(\frac{2M}{\pi RT_v}\right)^{1/2}\left[\frac{h_{s,v}}{\rho c_p}(\rho_v - \rho_\infty) + \frac{M}{5RT_v(0)}h_{s,v}[T_v(0) - T_\infty]\right] \tag{45}$$

For the liquid film, we can assume conduction only, which means

$$q_w = k_l\frac{T_w - T_l(0)}{d} \tag{46}$$

Equations (42), (43), (45) can be combined with two interfacial conditions for mass and heat flux in Sec. 3, and Eqs. (39) and (46) to solve for q, $\dot{m}$, $u_\infty$, $T_v(0)$, $\rho_v(0)$, $T_l(0)$ and $q_w$ (seven equations with seven unknowns), assuming $\rho_\infty$ and $T_\infty$ and $h_s$ are known. These equations are solved numerically. We will give some examples of the solution below.

Figure 4(a) shows the liquid and vapor phase temperatures as a function of the liquid film thickness and Fig. 4(b) the interfacial temperature differences. Figure 4(c) shows the overall heat transfer coefficients and Fig. 4(d) the mass flux for specific sets of given conditions that correspond to typical heat transfer applications. We can see that in these cases, the liquid side temperature is always higher than the vapor side, because the liquid film is thin and can supply sufficient heat for evaporation, leading to a vapor phase heat flux also in the same direction as the liquid phase. The higher is the single-phase heat transfer coefficient, the larger is the interface temperature jump and the higher is the overall heat transfer coefficient (Fig. 4(c)). Nonideal accommodation coefficient ($\alpha$ not equaling one) leads to lower overall heat transfer coefficient and mass evaporation rate. For an overall temperature drop of 20 $^{o}$C, the interfacial temperature drop could be 0.5-2 $^{o}$C. Most of the temperature drops occur in the vapor phase.

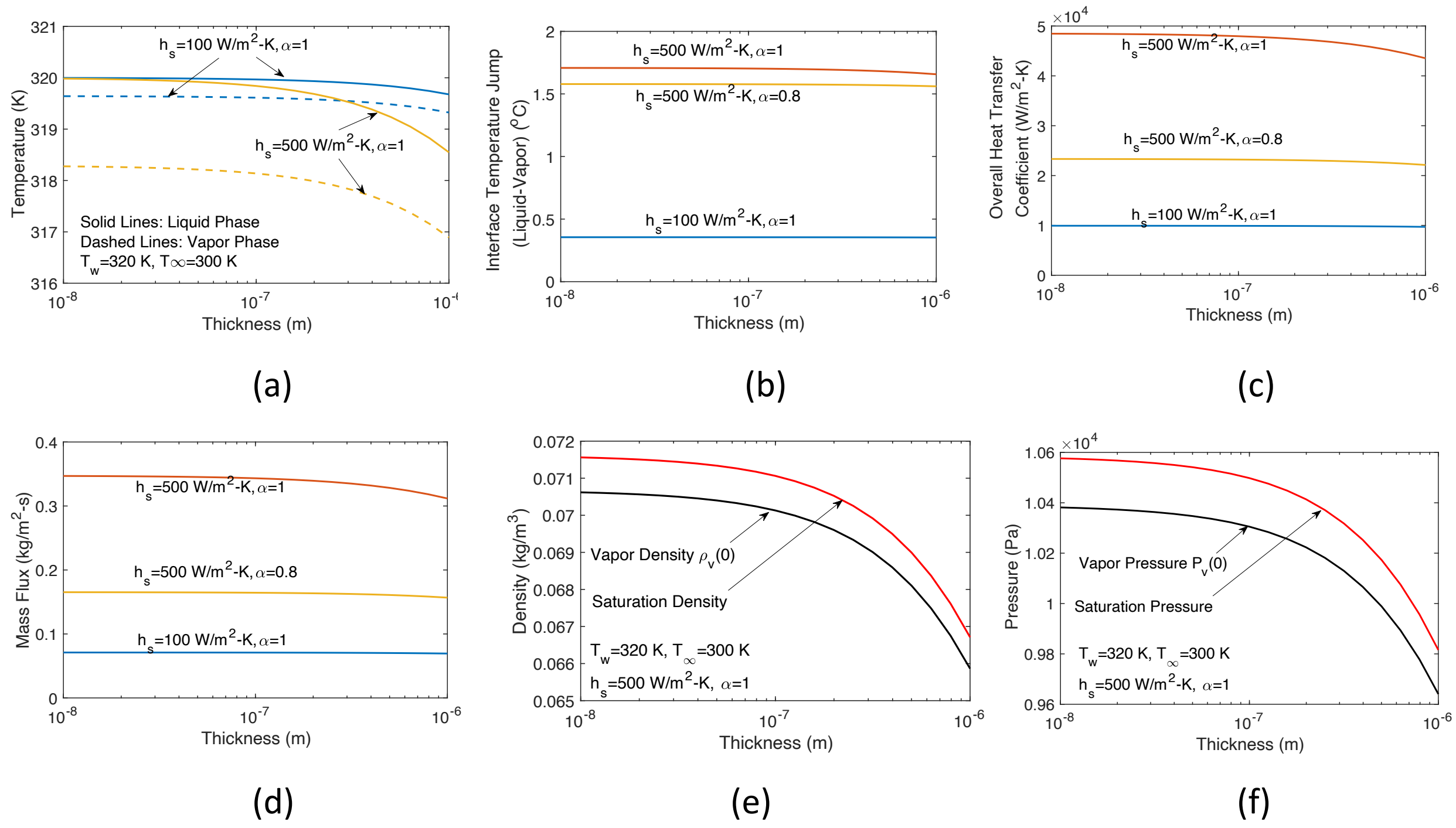

Figure 4. Simulation results for single interface evaporation heat transfer as a function of the liquid film thickness at different single phase heat transfer coefficient and accommodation coefficient values (a) the liquid and the vapor phase temperatures at the interface and (b) the temperature drops across the liquid-vapor interface, (c) overall heat transfer coefficient, (d) evaporation mass flux, (e) saturation density and vapor side density, and (f) saturation pressure and vapor side pressure at the interface. All for solid wall temperature $T_w$=320 K, and ambient $T_\infty$=300 K.

Accompanying the interface temperature discontinuity are also the density and pressure discontinuities [Fig.4(e) and 4(f)]. The saturation density and pressure, which are presumably the vapor properties right on the liquid surface, are higher than their corresponding values on the vapor side, i.e., at the outer edge of the Knudsen layer. At $10^4$ Pa, the mean free path is ~1 μm and the Knudsen layer is a few times of the mean free path, which we treated as zero thickness. Although the density and pressure differences are small, they are the driven force for the mass transport. In some sense, the temperature drop across the interface can be thought as the interfacial Joule-Thomson effect. The vapor side temperature is colder due to the sudden expansion from the liquid phase to the vapor phase.

## 5. Conclusion

In this work, we established interfacial conditions for evaporation and condensation at the interface based on the kinetic theory. These interfacial conditions are consistently derived from the solution of the BTE in the bulk region, based on the general diffusion-transmission boundary condition we established before for interfacial transport. Three interfacial boundary conditions are derived, one is for the mass flux, which is identical to the well-known Schrage equation. The other is for the heat flux. We also derive an additional interfacial condition for the density discontinuity across the interface. The three equations can be used to couple transport in the bulk liquid and vapor regions to determine the interfacial discontinuities in temperature, density, and pressure.

We show that our approach can well explain the experimental data reporting interfacial temperature discontinuities, while previous modeling efforts had failed to explain quantitatively. We also used these interfacial conditions to connect the evaporation heat transfer coefficient to the single-phase heat transfer coefficient between the vapor and the liquid phase. Our results show that the interfacial temperature drop is appreciable in typical heat transfer experiments.

In addition to the interfacial conditions, we also present a thermomolecular emission model for evaporation, mimicking the well-established thermionic emission theory for electrons. Although the final results can all be based on the vapor-phase properties, this model gives more physical insights on thermal evaporation. The evaporation and condensation coefficients distinguished in the past literature are likely due to the strong temperature dependence of the condensation coefficient. More work is needed to refine the model for better understanding of the molecular picture of phase-change processes.

We should point out that our work did not include the internal degree of freedom in the molecules. Past studies suggest that we can include them by, for example, replacing the 5R/2M in the enthalpy expression with the constant pressure specific heat. We did not include nonlinear effect, although the SM materials provided can be used to extend the formulation to include these effects. Also, the vapor-phase is pure substance. Further work to include the mixtures, especially non-condensable gases should be conducted.

**Acknowledgments.** I would like to thank MIT for its support during a difficult time of my life. I thank Mr. Buxuan Li for his help in coding and Mr. Simo Pajovic for proofreading the manuscript. I would like to thank Professors Hadi Ghasemi (U. Houston), John Lienhard (MIT), Yangying Zhu (UCSB) for their constructive comments. Special thanks also go to Dr. Zhengmao Lu and Mr. Geoffrey Vaartstra for stimulating discussions, especially with regarding to the mass flux expression, and sharing an unpublished manuscript.

# Supplemental Materials

# On the Molecular Picture and Interfacial Temperature Discontinuity During Evaporation and Condensation

Gang Chen

Department of Mechanical Engineering
Massachusetts Institute of Technology

## Sec. 1. Thermomolecular Emission Current

The outgoing molecular flux is obtained from

$$J_s^+ = \varepsilon \int_{-\infty}^{\infty} dv_x \int_{-\infty}^{\infty} dv_y \int_{\sqrt{2\Delta/m}}^{\infty} v_z f_L dv_z$$

$$= \varepsilon n_l \sqrt{\frac{k_B T_l}{2\pi m}} exp\left\{-\frac{\Delta}{k_B T_l}\right\} \int_{-\infty}^{\infty} dv_x \int_{-\infty}^{\infty} dv_y \int_{\sqrt{2\Delta/m}}^{\infty} v_z n_l \left[\frac{m}{2\pi k_B T_l}\right]^{3/2} exp\left\{-\frac{m[v_x^2+v_y^2+v_z^2]}{2k_B T_l}\right\} dv_z$$

$$= \varepsilon n_l \sqrt{\frac{k_B T_l}{2\pi m}} exp\left\{-\frac{\Delta}{k_B T_l}\right\} \qquad \text{(A1)}$$

In the above expression, we assume only molecules with kinetic energy component in the z-direction larger than the potential barrier can escape the liquid surface. Another extreme is molecules with total energy larger than the potential barrier can escape the liquid surface, i.e.,

$$\frac{m}{2}\left(v_x^2 + v_y^2 + v_z^2\right) > \Delta \qquad \text{and} \qquad v_z > 0 \qquad \text{(A2)}$$

Integration for the molecular flux leaving the surface with the above condition can be done by using spherical coordinate

$$J_s^+ = \varepsilon \int_0^{2\pi} d\varphi \int_0^{\pi/2} d\theta \int_{\sqrt{2\Delta/m}}^{\infty} v cos\theta \, n_l \left[\frac{m}{2\pi k_B T_l}\right]^{3/2} exp\left\{-\frac{mv^2}{2k_B T_l}\right\} sin\theta v^2 dv$$

$$= \varepsilon n_l \sqrt{\frac{k_B T_l}{2\pi m}} \left(1 + \frac{\Delta}{k_B T_l}\right) exp\left\{-\frac{\Delta}{k_B T_l}\right\} \qquad \text{(A3)}$$

This flux is much larger than (A1) since $\frac{\Delta}{k_B T_l}$ is ~10 for water. In comparison, the Richardson formula for electron thermionic emission is

$$J_s^+ = AT^2 exp\left\{-\frac{\Delta}{k_B T_l}\right\} \qquad \text{(A4)}$$

where A is the Richardson constant. The difference comes from the fact that electron statistical distribution, even in the limit that the Boltzmann statistics is valid, is different from molecules since electrons are Fermions. Richardson originally had same $T^{1/2}$ dependence in the prefactor similar to (A1) and (A3) because he used same Maxwell-Boltzmann distribution function.

**Sec. 2. Mass and Heat Fluxes in the Bulk Region**

The diffusion approximation is well discussed in standard textbooks (28,34). For completeness, we will give the derivation here. We will first write done details some useful integrals that we will use often. For such purpose, it is useful to start with the following integral

$$F(a)=\int_{-\infty}^{\infty} exp(-av_x^2)dv_x = \sqrt{\frac{\pi}{a}} \qquad \text{(A5)}$$

Setting $a = m/(2k_BT)$, we get

$$\int_{-\infty}^{\infty} exp\left(-\frac{mv_x^2}{2k_BT}\right) dv_x = \sqrt{\frac{2\pi k_BT}{m}} \qquad \text{(A6)}$$

$$\int_{-\infty}^{\infty} v_x^2 exp\left(-\frac{mv_x^2}{2k_BT}\right) dv_x = -\frac{dF(a)}{da}\Big)_{a=m/(2k_BT)} = \frac{\sqrt{\pi}}{2}\left(\frac{2k_BT}{m}\right)^{3/2} \qquad \text{(A7)}$$

$$\int_{-\infty}^{\infty} v_x^4 exp\left(-\frac{mv_x^2}{2k_BT}\right) dv_x = \frac{d^2F(a)}{da^2}\Big)_{a=m/(2k_BT)} = \frac{3\sqrt{\pi}}{4}\left(\frac{2k_BT}{m}\right)^{5/2} \qquad \text{(A8)}$$

$$\int_{-\infty}^{\infty} v_x^6 exp\left(-\frac{mv_x^2}{2k_BT}\right) dv_x = -\frac{d^3F(a)}{da^3}\Big)_{a=m/(2k_BT)} = \frac{15\sqrt{\pi}}{8}\left(\frac{2k_BT}{m}\right)^{7/2} \qquad \text{(A9)}$$

**Mass Flux.** To evaluate the molecular flux,

$$J_m(z) = \int_{-\infty}^{\infty} dv_x \int_{-\infty}^{\infty} dv_y \int_{-\infty}^{\infty} v_z f dv_z = un(z) -$$

$$-\int_{-\infty}^{\infty} dv_x \int_{-\infty}^{\infty} dv_y \int_{-\infty}^{\infty} f_d \tau v_z^2 \left\{\frac{1}{n}\frac{dn}{dz} + \frac{1}{T}\frac{dT}{dz}\left[\frac{m\left[v_x^2+v_y^2+(v_z-u(z))^2\ \right]}{2k_BT(z)} - \frac{3}{2}\right] + \frac{m[v_z-u(z)]}{k_BT(z)}\frac{du}{dz}\right\} dv_z \qquad \text{(A10)}$$

Next we evaluate each term in Eq. (A10):

$$\int_{-\infty}^{\infty} dv_x \int_{-\infty}^{\infty} dv_y \int_{-\infty}^{\infty} f_d \tau v_z^2 \left\{\frac{1}{n}\frac{dn}{dz}\right\} dv_z$$

$$= \tau n \left\{\frac{1}{n}\frac{dn}{dz}\right\} \left[\frac{m}{2\pi k_BT(z)}\right]^{3/2} \int_o^{\infty} 2\pi v_{//} exp\left[-\frac{mv_{//}^2}{2k_BT}\right] dv_{//} \int_{-\infty}^{\infty} v_z^2 expp\left[-\frac{m(v_z-u(z))^2}{2k_BT}\right] dv_z$$

$$= \tau n \left\{\frac{1}{n}\frac{dn}{dz}\right\} \left[\frac{m}{2\pi k_B T}\right]^{3/2} \pi \left(\frac{2k_B T}{m}\right) \int_{-\infty}^{\infty} \left[(v_z - u)^2 \; + 2u(v_z - u) + u^2\right] exp\left[-\frac{m(v_z-u)^2}{2k_B T}\right] dv_z$$

$$= \tau n \left\{\frac{1}{n}\frac{dn}{dz}\right\} \pi \left(\frac{2k_B T}{m}\right) \left[\frac{m}{2\pi k_B T}\right]^{3/2} \left[\frac{\sqrt{\pi}}{2}\left(\frac{2k_B T}{m}\right)^{3/2} + u^2\sqrt{\frac{2\pi k_B T}{m}}\right]$$

$$= \tau \frac{dn}{dz}\left(\frac{2k_B T}{m}\right)\left[\frac{1}{2} + \frac{mu^2}{2k_B T}\right]$$

$$\approx \tau \frac{dn}{dz}\left(\frac{k_B T}{m}\right) \qquad \text{(A11)}$$

Note that the odd order ($v_z$-u) term drops out because of symmetry in the integration. Later, when we consider the boundary condition, this term will not drop out. The last step kept only the lowest order term, assuming the average velocity u is much smaller than thermal velocity. This higher order term can be kept if the evaporation rate is very high.

Next, we consider dT/dz term in Eq. (A10)

$$\int_{-\infty}^{\infty} dv_x \int_{-\infty}^{\infty} dv_y \int_{-\infty}^{\infty} f_d \tau v_z^2 \left\{\frac{1}{T}\frac{dT}{dz}\left[\frac{m\left[v_x^2+v_y^2+(v_z-u(z))^2\right]}{2k_B T(z)} - \frac{3}{2}\right]\right\} dv_z$$

$$= n\tau \left[\frac{m}{2\pi k_B T}\right]^{3/2} \frac{1}{T}\frac{dT}{dz} \int_{-\infty}^{\infty} dv_x \int_{-\infty}^{\infty} dv_y \int_{-\infty}^{\infty} v_z^2 \left\{\left[\frac{m\left[v_x^2+v_y^2+(v_z-u)^2\right]}{2k_B T} - \frac{3}{2}\right]\right\} dv_z$$

$$= n\tau \left[\frac{m}{2\pi k_B T}\right]^{3/2} \frac{1}{T}\frac{dT}{dz}\left[\frac{m}{2k_B T}\int_o^{\infty} 2\pi v_{//}{}^3 exp\left(-\frac{mv_{//}^2}{2k_B T}\right) dv_{//} \int_{-\infty}^{\infty} v_z^2 exp\left(-\frac{m(v_z-u)^2}{2k_B T}\right) dv_z\right.$$

$$+ \frac{m}{2k_B T}\int_o^{\infty} 2\pi v_{//} \quad exp\left(-\frac{mv_{//}^2}{2k_B T}\right) dv_{//} \int_{-\infty}^{\infty} v_z^2 (v_z - u)^2 \; exp\left(-\frac{m(v_z-u)^2}{2k_B T}\right) dv_z$$

$$\left. - \frac{3}{2}\int_o^{\infty} 2\pi v_{//} \quad exp\left(-\frac{mv_{//}^2}{2k_B T}\right) dv_{//} \int_{-\infty}^{\infty} v_z^2 exp\left(-\frac{m(v_z-u)^2}{2k_B T}\right) dv_z\right]$$

$$= n\tau \left[\frac{m}{2\pi k_B T}\right]^{3/2} \frac{1}{T}\frac{dT}{dz}\left[\frac{m}{2k_B T}\pi\left(\frac{2k_B T}{m}\right)^2 \int_{-\infty}^{\infty}\left[v_z'^2 + 2uv_z' + u^2\right] exp\left(-\frac{mv_z'^2}{2k_B T}\right) dv'_z\right.$$

$$+ \frac{m}{2k_B T}\pi\left(\frac{2k_B T}{m}\right)\int_{-\infty}^{\infty}\left[v_z'^4 + 2uv_z'^3 + u^2 v_z'^2\right] exp\left(-\frac{mv_z'^2}{2k_B T}\right) dv'_z$$

$$\left. - \frac{3}{2}\pi\left(\frac{2k_B T}{m}\right)\int_{-\infty}^{\infty}\left[v_z'^2 + 2uv_z' + u^2\right] exp\left(-\frac{mv_z'^2}{2k_B T}\right) dv'_z\right]$$

$$= n\tau\left[\frac{m}{2\pi k_B T}\right]^{3/2}\frac{1}{T}\frac{dT}{dz}\left[\frac{m}{2k_B T}\pi\left(\frac{2k_B T}{m}\right)^2\langle\frac{\sqrt{\pi}}{2}\left(\frac{2k_B T}{m}\right)^{3/2}+u^2\sqrt{\frac{2\pi k_B T}{m}}\rangle\right.$$

$$\left.+\frac{m}{2k_B T}\pi\left(\frac{2k_B T}{m}\right)\langle\frac{3\sqrt{\pi}}{4}\left(\frac{2k_B T}{m}\right)^{5/2}+u^2\frac{\sqrt{\pi}}{2}\left(\frac{2k_B T}{m}\right)^{3/2}\rangle-\frac{3}{2}\pi\left(\frac{2k_B T}{m}\right)\langle\frac{\sqrt{\pi}}{2}\left(\frac{2k_B T}{m}\right)^{3/2}+u^2\sqrt{\frac{2\pi k_B T}{m}}\rangle\right]$$

$$= n\tau\frac{1}{T}\frac{dT}{dz}\left[\left(\frac{2k_B T}{m}\right)\left(\frac{1}{2}+\frac{mu^2}{2k_B T}\right)+\langle\frac{3}{4}\left(\frac{2k_B T}{m}\right)+\frac{1}{2}u^2\rangle-\frac{3}{2}\left(\frac{2k_B T}{m}\right)\left(\frac{1}{2}+\frac{mu^2}{2k_B T}\right)\right]$$

$$= n\tau\frac{1}{T}\frac{dT}{dz}\left(\frac{k_B T}{m}\right) \tag{A12}$$

Next, we evaluate the du/dz term

$$\int_{-\infty}^{\infty}dv_x\int_{-\infty}^{\infty}dv_y\int_{-\infty}^{\infty}f_d\tau v_z^2\left\{\frac{m[v_z-u(z)]}{k_B T(z)}\frac{du}{dz}\right\}dv_z$$

$$= \tau n\frac{m}{k_B T}\frac{du}{dz}\int_o^{\infty}2\pi v_{//}\ \ exp\left(-\frac{mv_{//}^2}{2k_B T}\right)dv_{//}\int_{-\infty}^{\infty}(v_z-u)v_z^2 exp\left(-\frac{m(v_z-u)^2}{2k_B T}\right)dv_z$$

$$= \tau n\frac{m}{k_B T}\left[\frac{m}{2\pi k_B T}\right]^{3/2}\frac{du}{dz}\int_o^{\infty}2\pi v_{//}\ \ exp\left(-\frac{mv_{//}^2}{2k_B T}\right)dv_{//}\int_{-\infty}^{\infty}(v_z-u)v_z^2 exp\left(-\frac{m(v_z-u)^2}{2k_B T}\right)dv_z$$

$$= \tau n\frac{m}{k_B T}\left[\frac{m}{2\pi k_B T}\right]^{3/2}\frac{du}{dz}\pi\left(\frac{2k_B T}{m}\right)2u\frac{\sqrt{\pi}}{2}\left(\frac{2k_B T}{m}\right)^{3/2}$$

$$= 2\tau nu\frac{du}{dz} \tag{A13}$$

Combining all terms, the molecular flux can be expressed as

$$J_m(z) = nu-\tau\frac{dn}{dz}\left(\frac{2k_B T}{m}\right)\left[\frac{1}{2}+\frac{mu^2}{2k_B T}\right]-n\tau\frac{1}{T}\frac{dT}{dz}\left(\frac{k_B T}{m}\right)-2\tau nu\frac{du}{dz}$$

$$\approx nu-\frac{k_B T}{m}\tau\frac{dn}{dz}-\frac{k_B T}{m}n\tau\frac{1}{T}\frac{dT}{dz} \tag{A14}$$

Multiplying the above expression by individual molecule's mass, we obtain the mass flux expression in Eq. (13).

**Heat Flux.** Next, we consider the heat flux,

$$q(z) = \int_{-\infty}^{\infty}dv_x\int_{-\infty}^{\infty}dv_y\int_{-\infty}^{\infty}v_z\frac{m}{2}\left[v_x^2+v_y^2+v_z^2\right]f dv_z \tag{A15}$$

Substitute f in Eq. (12) into the above expression, we again see that the energy flux has convection term, the density, temperature and velocity gradient terms, as below.

Convection term

$$\int_{-\infty}^{\infty} dv_x \int_{-\infty}^{\infty} dv_y \int_{-\infty}^{\infty} v_z \frac{m}{2}\left[v_x^2 + v_y^2 + v_z^2\right] f_d dv_z$$

$$= \frac{m}{2} n \left[\frac{m}{2\pi k_B T}\right]^{3/2} \left[\int_o^{\infty} 2\pi v_{//}{}^3 exp\left(-\frac{mv_{//}^2}{2k_B T}\right) dv_{//} \int_{-\infty}^{\infty} v_z exp\left(-\frac{m(v_z-u)^2}{2k_B T}\right) dv_z\right.$$

$$\left. + \int_o^{\infty} 2\pi v_{//} \;\; exp\left(-\frac{mv_{//}^2}{2k_B T}\right) dv_{//} \int_{-\infty}^{\infty} v_z{}^3 exp\left(-\frac{m(v_z-u)^2}{2k_B T}\right) dv_z\right]$$

$$= \frac{m}{2} n \left[\frac{m}{2\pi k_B T}\right]^{3/2} \left[\pi \left(\frac{2k_B T}{m}\right)^2 \int_{-\infty}^{\infty} [v'_z + u] exp\left(-\frac{mv_z'^2}{2k_B T}\right) dv_z\right.$$

$$\left. + \pi \left(\frac{2k_B T}{m}\right) \;\; \int_{-\infty}^{\infty} [v_z{}'^3 + 3uv_z{}'^2 + 3u^2 v_z{}' + u^3] exp\left(-\frac{mv_z'^2}{2k_B T}\right) dv_z\right]$$

$$= \frac{mn}{2}\left(\frac{2k_B T}{m}\right) u \;+ \frac{mn}{2}\left(\frac{2k_B T}{m}\right) \;\; u\left[\frac{3}{2} + \frac{mu^2}{2k_B T}\right]$$

$$= \frac{mn}{2}\left(\frac{2k_B T}{m}\right) \;\; u\left[\frac{5}{2} + \frac{mu^2}{2k_B T}\right] \qquad \text{(A16)}$$

The mass diffusion term dn/dz is as follows

$$\int_{-\infty}^{\infty} dv_x \int_{-\infty}^{\infty} dv_y \int_{-\infty}^{\infty} f_d \tau v_z^2 \frac{m}{2}\left[v_x^2 + v_y^2 + v_z^2\right]\left\{\frac{1}{n}\frac{dn}{dz}\right\} dv_z$$

$$= \frac{m}{2} n\tau \left[\frac{m}{2\pi k_B T}\right]^{3/2} \frac{1}{n}\frac{dn}{dz}\left[\int_o^{\infty} 2\pi v_{//}{}^3 exp\left(-\frac{mv_{//}^2}{2k_B T}\right) dv_{//} \int_{-\infty}^{\infty} v_z{}^2 exp\left(-\frac{m(v_z-u)^2}{2k_B T}\right) dv_z\right.$$

$$\left. + \int_o^{\infty} 2\pi v_{//} \;\; exp\left(-\frac{mv_{//}^2}{2k_B T}\right) dv_{//} \int_{-\infty}^{\infty} v_z{}^4 exp\left(-\frac{m(v_z-u)^2}{2k_B T}\right) dv_z\right]$$

$$= \frac{m}{2} n\tau \left[\frac{m}{2\pi k_B T}\right]^{3/2} \frac{1}{n}\frac{dn}{dz}\left[\pi \left(\frac{2k_B T}{m}\right)^2 \int_{-\infty}^{\infty} [v_z{}'^2 + 2uv_z{}' + u^2] exp\left(-\frac{mv_z{}'^2}{2k_B T}\right) dv_z{}'\right.$$

$$\left. + \pi \left(\frac{2k_B T}{m}\right) \;\; \int_{-\infty}^{\infty} [v_z{}'^4 + 6u^2 v_z'^2 + u^4] exp\left(-\frac{mv_z{}'^2}{2k_B T}\right) dv_z{}'\right]$$

$$= \frac{m}{2} n\tau \left[\frac{m}{2\pi k_B T}\right]^{3/2} \frac{1}{n}\frac{dn}{dz}\left[\pi \left(\frac{2k_B T}{m}\right)^2 \langle \frac{\sqrt{\pi}}{2}\left(\frac{2k_B T}{m}\right)^{3/2} + u^2\sqrt{\frac{2\pi k_B T}{m}}\rangle\right.$$

$$\left. + \pi \left(\frac{2k_B T}{m}\right) \;\; \langle \frac{3\sqrt{\pi}}{4}\left(\frac{2k_B T}{m}\right)^{5/2} + 6u^2 \frac{\sqrt{\pi}}{2}\left(\frac{2k_B T}{m}\right)^{3/2} + u^4\sqrt{\frac{2\pi k_B T}{m}}\rangle\right]$$

$$= \frac{m}{2} n\tau \frac{1}{n}\frac{dn}{dz}\left[\left(\frac{2k_BT}{m}\right)^2\left(\frac{1}{2}+\frac{mu^2}{2k_BT}\right)+\left(\frac{2k_BT}{m}\right)^2\left(\frac{3}{4}+\frac{3mu^2}{2k_BT}+\left[\frac{mu^2}{2k_BT}\right]^2\right)\right]$$

$$= \frac{m}{2} n\tau \left(\frac{2k_BT}{m}\right)^2\left(\frac{5}{4}+\frac{2mu^2}{k_BT}+\left[\frac{mu^2}{2k_BT}\right]^2\right)\frac{1}{n}\frac{dn}{dz} \qquad \text{(A17)}$$

The temperature gradient term dT/dz is:

$$\int_{-\infty}^{\infty} dv_x \int_{-\infty}^{\infty} dv_y \int_{-\infty}^{\infty} f_d \tau v_z^2 \frac{m}{2}\left[v_x^2+v_y^2+v_z^2\right]\frac{1}{T}\frac{dT}{dz}\left[\frac{m\left[v_x^2+v_y^2+(v_z-u(z))^2\right]}{2k_BT(z)}-\frac{3}{2}\right]dv_z$$

$$= \frac{m}{2} n\tau \left[\frac{m}{2\pi k_BT}\right]^{3/2}\frac{1}{T}\frac{dT}{dz}\left[\int_o^{\infty} 2\pi v_{//} exp\left(-\frac{mv_{//}^2}{2k_BT}\right)dv_{//} \times\right.$$

$$\left.\int_{-\infty}^{\infty} v_z^2\left(v_{//}^2+v_z^2\right)\langle\frac{m\left[v_{//}^2+(v_z-u)^2\right]}{2k_BT}-\frac{3}{2}\rangle\, exp\left(-\frac{m(v_z-u)^2}{2k_BT}\right)dv_z\right.$$

$$= \frac{m}{2} n\tau \left[\frac{m}{2\pi k_BT}\right]^{3/2}\frac{1}{T}\frac{dT}{dz}\times\left[\frac{m}{2k_BT}\langle\int_o^{\infty} 2\pi v_{//}{}^5 exp\left(-\frac{mv_{//}^2}{2k_BT}\right)dv_{//}\int_{-\infty}^{\infty} v_z^2 exp\left(-\frac{m(v_z-u)^2}{2k_BT}\right)dv_z\rangle\right.$$

$$+\frac{m}{2k_BT}\langle\int_o^{\infty} 2\pi v_{//}{}^3 exp\left(-\frac{mv_{//}^2}{2k_BT}\right)dv_{//}\int_{-\infty}^{\infty}\left[v_z^4+v_z^2(v_z-u)^2\right]exp\left(-\frac{m(v_z-u)^2}{2k_BT}\right)dv_z\rangle$$

$$+\frac{m}{2k_BT}\langle\int_o^{\infty} 2\pi v_{//}\ \ exp\left(-\frac{mv_{//}^2}{2k_BT}\right)dv_{//}\int_{-\infty}^{\infty} v_z^4(v_z-u)^2\ exp\left(-\frac{m(v_z-u)^2}{2k_BT}\right)dv_z\rangle$$

$$-\frac{3}{2}\langle\int_o^{\infty} 2\pi v_{//}{}^3 exp\left(-\frac{mv_{//}^2}{2k_BT}\right)dv_{//}\int_{-\infty}^{\infty} v_z^2 exp\left(-\frac{m(v_z-u)^2}{2k_BT}\right)dv_z\rangle$$

$$\left.-\frac{3}{2}\int_o^{\infty} 2\pi v_{//}\ \ exp\left(-\frac{mv_{//}^2}{2k_BT}\right)dv_{//}\int_{-\infty}^{\infty} v_z^4 exp\left(-\frac{m(v_z-u)^2}{2k_BT}\right)dv_z\right]$$

$$= \frac{m}{2} n\tau \left[\frac{m}{2\pi k_BT}\right]^{3/2}\frac{1}{T}\frac{dT}{dz}\times\left[\frac{m}{2k_BT}2\pi\left(\frac{2k_BT}{m}\right)^3\langle\int_{-\infty}^{\infty}\left[v_z'^2+2uv_z+u^2\right]exp\left(-\frac{mv_z'^2}{2k_BT}\right)dv_z\rangle\right.$$

$$+\frac{m}{2k_BT}\pi\left(\frac{2k_BT}{m}\right)^2\langle\int_{-\infty}^{\infty}\left[2v_z'^4+3u^2v_z'^2+2uv_z'^3+u^4\right]exp\left(-\frac{mv_z'^2}{2k_BT}\right)dv'_z\rangle$$

$$+\frac{m}{2k_BT}\pi\left(\frac{2k_BT}{m}\right)\quad\langle\int_{-\infty}^{\infty}\left[v_z'^6+2u^2v_z'^4+u^4v_z'^2\right]exp\left(-\frac{mv_z'^2}{2k_BT}\right)dv_z\rangle$$

$$-\frac{3}{2}\pi\left(\frac{2k_BT}{m}\right)^2\langle\int_{-\infty}^{\infty}\left[v_z'^2+2uv'_z+u^2\right]exp\left(-\frac{mv_z'^2}{2k_BT}\right)dv_z\rangle$$

$$-\frac{3}{2}\pi\left(\frac{2k_BT}{m}\right) \quad \int_{-\infty}^{\infty}\left[v_z'^4+2u^4\,{v_z}'^2+u^4\right]exp\left(-\frac{m(v_z-u)^2}{2k_BT}\right)dv_z\Bigg]$$

$$=\frac{m}{2}n\tau\left[\frac{m}{2\pi k_BT}\right]^{3/2}\frac{1}{T}\frac{dT}{dz}\times\Bigg[\frac{m}{2k_BT}2\pi\left(\frac{2k_BT}{m}\right)^3\langle\frac{\sqrt{\pi}}{2}\left(\frac{2k_BT}{m}\right)^{3/2}+u^2\sqrt{\frac{2\pi k_BT}{m}}\rangle$$

$$+\frac{m}{2k_BT}\pi\left(\frac{2k_BT}{m}\right)^2\langle 2\,\frac{3\sqrt{\pi}}{4}\left(\frac{2k_BT}{m}\right)^{5/2}+3u^2\,\frac{\sqrt{\pi}}{2}\left(\frac{2k_BT}{m}\right)^{3/2}+u^4\sqrt{\frac{2\pi k_BT}{m}}\rangle$$

$$+\frac{m}{2k_BT}\pi\left(\frac{2k_BT}{m}\right) \quad \langle\frac{15\sqrt{\pi}}{8}\left(\frac{2k_BT}{m}\right)^{7/2}+2u^2\,\frac{3\sqrt{\pi}}{4}\left(\frac{2k_BT}{m}\right)^{5/2}+u^4\,\frac{\sqrt{\pi}}{2}\left(\frac{2k_BT}{m}\right)^{3/2}\rangle$$

$$-\frac{3}{2}\pi\left(\frac{2k_BT}{m}\right)^2\langle\frac{\sqrt{\pi}}{2}\left(\frac{2k_BT}{m}\right)^{3/2}+u^2\sqrt{\frac{2\pi k_BT}{m}}\rangle$$

$$-\frac{3}{2}\pi\left(\frac{2k_BT}{m}\right) \quad \langle\frac{3\sqrt{\pi}}{4}\left(\frac{2k_BT}{m}\right)^{5/2}+u^2\,\frac{\sqrt{\pi}}{2}\left(\frac{2k_BT}{m}\right)^{3/2}+u^4\sqrt{\frac{2\pi k_BT}{m}}\rangle\Bigg]$$

$$=\frac{m}{2}n\tau\frac{1}{T}\frac{dT}{dz}\times\Bigg[\frac{m}{2k_BT}2\left(\frac{2k_BT}{m}\right)^3\langle\frac{1}{2}+\frac{mu^2}{2k_BT}\rangle+\frac{m}{2k_BT}\left(\frac{2k_BT}{m}\right)^2\langle 2\,\frac{3}{4}\left(\frac{2k_BT}{m}\right) \quad +3u^2\,\frac{1}{2}+\frac{mu^4}{2k_BT}\rangle$$

$$+\langle\frac{15}{8}\left(\frac{2k_BT}{m}\right)^2+2u^2\,\frac{3}{4}\left(\frac{2k_BT}{m}\right) \quad +u^4\,\frac{1}{2}\rangle-\frac{3}{2}\left(\frac{2k_BT}{m}\right)^2\langle\frac{1}{2}+\frac{mu^2}{2k_BT}\rangle$$

$$-\frac{3}{2}\left(\frac{2k_BT}{m}\right) \quad \langle\frac{3}{4}\left(\frac{2k_BT}{m}\right)^1+u^2\,\frac{1}{2}+\frac{mu^4}{2k_BT}\rangle\Bigg]$$

$$=\frac{m}{2}n\tau\frac{1}{T}\frac{dT}{dz}\times\left(\frac{2k_BT}{m}\right)^2\left(\frac{5}{2}+2\,\frac{mu^2}{2k_BT}\right) \qquad \text{(A18)}$$

The du/dz term is

$$\int_{-\infty}^{\infty}dv_x\int_{-\infty}^{\infty}dv_y\int_{-\infty}^{\infty}f_d\tau v_z^2\,\frac{m}{2}\left[v_x^2+v_y^2+v_z^2\right]\frac{m[v_z-u(z)]}{k_BT(z)}\frac{du}{dz}dv_z$$
$$=\frac{m}{2}\tau n\frac{m}{k_BT}\frac{du}{dz}\left[\frac{m}{2\pi k_BT}\right]^{3/2}\left[\int_o^{\infty}2\pi {v_{//}}^3 exp\left(-\frac{mv_{//}^2}{2k_BT}\right)dv_{//}\int_{-\infty}^{\infty}v_z^2[v_z-u(z)]exp\left(-\frac{m(v_z-u)^2}{2k_BT}\right)dv_z\right.$$

$$\left.+\int_o^{\infty}2\pi v_{//} \quad exp\left(-\frac{mv_{//}^2}{2k_BT}\right)dv_{//}\int_{-\infty}^{\infty}v_z^4[v_z-u(z)]exp\left(-\frac{m(v_z-u)^2}{2k_BT}\right)dv_z\right]$$

$$=\frac{m}{2}\tau n\frac{m}{k_BT}\frac{du}{dz}\left[\frac{m}{2\pi k_BT}\right]^{3/2}\left[\pi\left(\frac{2k_BT}{m}\right)^2\int_{-\infty}^{\infty} \quad \left[v_z'^3+2u{v_z}'^2+u^2{v_z}'\right]exp\left(-\frac{m{v_z}'^2}{2k_BT}\right)dv_z\right.$$

$$\left.+\pi\left(\frac{2k_BT}{m}\right) \quad \int_{-\infty}^{\infty}\left[v_z'^5+2u^2{v_z}'^3+u^4{v_z}'\right]exp\left(-\frac{m{v_z}'^2}{2k_BT}\right)d{v_z}'\right]$$

$$= \frac{m}{2}\tau n \frac{m}{k_BT}\frac{du}{dz}\left[\frac{m}{2\pi k_BT}\right]^{3/2} \pi\left(\frac{2k_BT}{m}\right)^2 2u\frac{\sqrt{\pi}}{2}\left(\frac{2k_BT}{m}\right)^{3/2}$$

$$= 2k_BT\tau nu\frac{du}{dz} \tag{A19}$$

Substituting each of the above terms into heat flux expression Eq. (A16), and keep only the lowest order terms, we obtain Eq. (14).

**Prove that ρu=constant in bulk region.** The density at any point can be written as

$$n(z) = \int_{-\infty}^{\infty} dv_x \int_{-\infty}^{\infty} dv_y \int_{-\infty}^{\infty} f\, dv_z = \int_{-\infty}^{\infty} dv_x \int_{-\infty}^{\infty} dv_y \int_{-\infty}^{\infty} f_d dv_z$$

$$- \int_{-\infty}^{\infty} dv_x \int_{-\infty}^{\infty} dv_y \int_{-\infty}^{\infty} f_d \tau v_z \left\{\frac{1}{n}\frac{dn}{dz} + \frac{1}{T}\frac{dT}{dz}\left[\frac{m[v_x^2+v_y^2+(v_z-u(z))^2]}{2k_BT(z)} - \frac{3}{2}\right] + \frac{m[v_z-u(z)]}{k_BT(z)}\frac{du}{dz}\right\} dv_z$$

The first the term related to $f_d$ is

$$\int_{-\infty}^{\infty} dv_x \int_{-\infty}^{\infty} dv_y \int_{-\infty}^{\infty} f_d dv_z = n_v(z)$$

The density gradient term:

$$\int_{-\infty}^{\infty} dv_x \int_{-\infty}^{\infty} dv_y \int_{-\infty}^{\infty} f_d \tau v_z \left\{\frac{1}{n}\frac{dn}{dz}\right\} dv_z$$

$$= \tau n\left\{\frac{1}{n}\frac{dn}{dz}\right\}\left[\frac{m}{2\pi k_BT(z)}\right]^{3/2} \int_o^{\infty} 2\pi v_{//} exp\left[-\frac{mv_{//}^2}{2k_BT}\right] dv_{//} \int_{-\infty}^{\infty} v_z\ exp\left[-\frac{m(v_z-u)^2}{2k_BT}\right] dv_z$$

$$= \tau n\left\{\frac{1}{n}\frac{dn}{dz}\right\}\left[\frac{m}{2\pi k_BT}\right]^{3/2} \pi\left(\frac{2k_BT}{m}\right) \int_{-\infty}^{\infty}[(v_z-u)+u] exp\left[-\frac{m(v_z-u)^2}{2k_BT}\right] dv_z$$

$$= \tau n\left\{\frac{1}{n}\frac{dn}{dz}\right\}\left[\frac{m}{2\pi k_BT}\right]^{3/2} \pi\left(\frac{2k_BT}{m}\right)\left[u\int_{-\infty}^{\infty} exp\left(-\frac{mv_z'^2}{2k_BT}\right) dv'_z\right]$$

$$= \tau n\left\{\frac{1}{n}\frac{dn}{dz}\right\}\pi\left(\frac{2k_BT}{m}\right)\left[\frac{m}{2\pi k_BT}\right]^{3/2}\left[u\sqrt{\frac{2\pi k_BT}{m}}\right]$$

$$= \tau u\ \frac{dn}{dz}$$

The dT/dz term is

$$\int_{-\infty}^{\infty} dv_x \int_{-\infty}^{\infty} dv_y \int_{-\infty}^{\infty} f_d \tau v_z \ \left\{\frac{1}{T}\frac{dT}{dz}\left[\frac{m\left[v_x^2+v_y^2+(v_z-u(z))^2\right]}{2k_B T(z)} - \frac{3}{2}\right]\right\} dv_z$$

$$= n\tau \left[\frac{m}{2\pi k_B T}\right]^{3/2} \frac{1}{T}\frac{dT}{dz} \int_{-\infty}^{\infty} dv_x \int_{-\infty}^{\infty} dv_y \int_{-\infty}^{\infty} v_z \ \left[\frac{m\left[v_x^2+v_y^2+(v_z-u)^2\right]}{2k_B T} - \frac{3}{2}\right] f_d dv_z$$

$$= n\tau \left[\frac{m}{2\pi k_B T}\right]^{3/2} \frac{1}{T}\frac{dT}{dz} \left[\frac{m}{2k_B T}\int_o^{\infty} 2\pi {v_{//}}^3 exp\left(-\frac{mv_{//}^2}{2k_B T}\right) dv_{//} \int_{-\infty}^{\infty} v_z \ exp\left(-\frac{m(v_z-u)^2}{2k_B T}\right) dv_z\right.$$

$$+ \frac{m}{2k_B T}\int_o^{\infty} 2\pi v_{//} \ \ exp\left(-\frac{mv_{//}^2}{2k_B T}\right) dv_{//} \int_{-\infty}^{\infty} v_z \ (v_z-u)^2 \ exp\left(-\frac{m(v_z-u)^2}{2k_B T}\right) dv_z$$

$$\left.- \frac{3}{2}\int_o^{\infty} 2\pi v_{//} \ \ exp\left(-\frac{mv_{//}^2}{2k_B T}\right) dv_{//} \int_{-\infty}^{\infty} v_z \ exp\left(-\frac{m(v_z-u)^2}{2k_B T}\right) dv_z\right]$$

$$= n\tau \left[\frac{m}{2\pi k_B T}\right]^{3/2} \frac{1}{T}\frac{dT}{dz} \left[\frac{m}{2k_B T}\pi\left(\frac{2k_B T}{m}\right)^2 \int_{-\infty}^{\infty} [v_z' + u] exp\left(-\frac{mv_z'^2}{2k_B T}\right) dv'_z\right.$$

$$\left.+ \frac{m}{2k_B T}\pi\left(\frac{2k_B T}{m}\right) \int_{-\infty}^{\infty} [v_z'^3 + uv_z'^2] exp\left(-\frac{mv_z'^2}{2k_B T}\right) dv'_z - \frac{3}{2}\pi\left(\frac{2k_B T}{m}\right)\int_{-\infty}^{\infty} [v_z' + u] exp\left(-\frac{mv_z'^2}{2k_B T}\right) dv'_z\right]$$

$$= n\tau \left[\frac{m}{2\pi k_B T}\right]^{3/2} \frac{1}{T}\frac{dT}{dz} \left[-\frac{\pi}{2}\left(\frac{2k_B T}{m}\right) \quad u\int_{-\infty}^{\infty} exp\left(-\frac{mv_z'^2}{2k_B T}\right) dv'_z + \pi u \int_{-\infty}^{\infty} v_z'^2 exp\left(-\frac{mv_z'^2}{2k_B T}\right) dv'_z\right]$$

$$= n\tau \left[\frac{m}{2\pi k_B T}\right]^{3/2} \frac{1}{T}\frac{dT}{dz} \left[-\frac{1}{2}u\left(\frac{2\pi k_B T}{m}\right)^{3/2} + \frac{1}{2}u\left(\frac{2\pi k_B T}{m}\right)^{3/2}\right]$$

$$= 0$$

The du/dz term is

$$\int_{-\infty}^{\infty} dv_x \int_{-\infty}^{\infty} dv_y \int_{-\infty}^{\infty} f_d \tau v_z \ \left\{\frac{m[v_z-u(z)]}{k_B T(z)}\frac{du}{dz}\right\} dv_z$$

$$= \tau n \frac{m}{k_B T}\left[\frac{m}{2\pi k_B T}\right]^{3/2} \frac{du}{dz}\int_o^{\infty} 2\pi v_{//} \ \ exp\left(-\frac{mv_{//}^2}{2k_B T}\right) dv_{//} \int_{-\infty}^{\infty} (v_z-u) v_z \ exp\left(-\frac{m(v_z-u)^2}{2k_B T}\right) dv_z$$

$$= \tau n \frac{m}{k_B T}\left[\frac{m}{2\pi k_B T}\right]^{3/2} \frac{du}{dz}\pi\left(\frac{2k_B T}{m}\right)\int_{-\infty}^{\infty} [(v_z-u)^2 + u(v_z-u)] exp\left(-\frac{m(v_z-u)^2}{2k_B T}\right) dv_z$$

$$= \tau n \frac{m}{k_B T}\left[\frac{m}{2\pi k_B T}\right]^{3/2} \frac{du}{dz}\pi\left(\frac{2k_B T}{m}\right)\int_{-\infty}^{\infty} [v_z'^2 + uv_z'] exp\left(-\frac{mv_z'^2}{2k_B T}\right) dv_z'$$

$$= \tau n \frac{m}{k_B T}\left[\frac{m}{2\pi k_B T}\right]^{3/2} \frac{du}{dz}\pi\left(\frac{2k_B T}{m}\right)\frac{\sqrt{\pi}}{2}\left(\frac{2k_B T}{m}\right)^{3/2}$$

$= \tau n \frac{du}{dz}$

So, the local density is

$$n(z) = n_v(z) + \tau u \ \frac{dn}{dz} + \tau n \frac{du}{dz} = n_v(z) + \tau \ \frac{d(un)}{dz} \qquad \text{(A20)}$$

For the above equation to be valid, nu=constant, i.e., ρu=constant.

### Sec. 3. Heat and Mass Fluxes at Interfaces

**Momentum Flux Coming Towards Interface.** Now, we consider the imbalance at the interface. Let us first consider the flux coming towards to interface, using the molecular distribution function again by Eq. (12). However, in this case, the integration of $v_z$ is for all values less than zero. The momentum flux is

$$J_m^-(z=0) = \int_{-\infty}^{\infty} dv_x \int_{-\infty}^{\infty} dv_y \int_{-\infty}^{o} v_z f_d dv_z$$

$$- \int_{-\infty}^{\infty} dv_x \int_{-\infty}^{\infty} dv_y \int_{-\infty}^{0} f_d \tau v_z^2 \left\{ \frac{1}{n}\frac{dn}{dz} + \frac{1}{T}\frac{dT}{dz}\left[\frac{m\left[v_x^2+v_y^2+(v_z-u(z))^2\ \right]}{2k_BT(z)} - \frac{3}{2}\right] + \frac{m[v_z-u(z)]}{k_BT(z)}\frac{du}{dz} \right\} dv_z \quad \text{(A21)}$$

We evaluate the first the term related to $f_d$

$$\int_{-\infty}^{\infty} dv_x \int_{-\infty}^{\infty} dv_y \int_{-\infty}^{0} f_d v_z dv_z = n\left[\frac{m}{2\pi k_BT}\right]^{3/2}$$

$$= n\left[\frac{m}{2\pi k_BT}\right]^{3/2} \int_{o}^{\infty} 2\pi v_{//} \ \ exp\left(-\frac{mv_{//}^2}{2k_BT}\right) dv_{//} \int_{-\infty}^{0} v_z exp\left(-\frac{m(v_z-u)^2}{2k_BT}\right) dv_z$$

$$= n\left[\frac{m}{2\pi k_BT}\right]^{3/2} \pi\left(\frac{2k_BT}{m}\right) \ \ \int_{-\infty}^{0} v_z exp\left(-\frac{m(v_z-u)^2}{2k_BT}\right) dv_z$$

$$= n\left[\frac{m}{2\pi k_BT}\right]^{3/2} \pi\left(\frac{2k_BT}{m}\right) \ \ \left[\int_{-\infty}^{0} [v_z-u) + u] exp\left(-\frac{m(v_z-u)^2}{2k_BT}\right) dv_z\right]$$

$$= n\left[\frac{m}{2\pi k_BT}\right]^{3/2} \pi\left(\frac{2k_BT}{m}\right) \ \ \left[\int_{-\infty}^{-u} v_z' exp\left(-\frac{mv_z'^2}{2k_BT}\right) dv'_z + u\int_{-\infty}^{-u} exp\left(-\frac{mv_z'^2}{2k_BT}\right) dv'_z\right]$$

$$= n\left[\frac{m}{2\pi k_BT}\right]^{3/2} \pi\left(\frac{2k_BT}{m}\right) \ \ \left[-\int_{u}^{\infty} v_z' exp\left(-\frac{mv_z'^2}{2k_BT}\right) dv'_z + u\int_{u}^{\infty} exp\left(-\frac{mv_z'^2}{2k_BT}\right) dv'_z\right]$$

$$= n\left[\frac{m}{2\pi k_B T}\right]^{1/2}\left[-\left(\frac{k_B T}{m}\right) \quad exp\left(-\frac{mu^2}{2k_B T}\right) + u\int_u^\infty exp\left(-\frac{m{v_z}^2}{2k_B T}\right)dv_z\right]$$

$$= n\left[\frac{m}{2\pi k_B T}\right]^{1/2}\left[-\left(\frac{k_B T}{m}\right) \quad exp\left(-\frac{mu^2}{2k_B T}\right) + u\sqrt{\frac{2k_B T}{m}}\,\frac{\sqrt{\pi}}{2}efrc\left(\sqrt{\frac{mu^2}{2k_B T}}\right)\right]$$

$$= -n\left[\left(\frac{k_B T}{2\pi m}\right)^{1/2} exp\left(-\frac{mu^2}{2k_B T}\right) - \frac{u}{2}erfc\left(\sqrt{\frac{mu^2}{2k_B T}}\right)\right]$$

$$= -n\left[\left(\frac{k_B T}{2\pi m}\right)^{1/2} exp(-\delta^2) - \frac{u}{2}erfc(\delta)\right] \tag{A22}$$

where

$$\delta = \sqrt{\frac{mu^2}{2k_B T}} \tag{A23}$$

and complementary error function

$$erfc(\delta) = \frac{2}{\sqrt{\pi}}\int_\delta^\infty exp(-x^2)dx \tag{A24}$$

Since u is typically much smaller than thermal velocity, (A22) can be approximated as

$$= -n\left[\left(\frac{k_B T}{2\pi m}\right)^{1/2} - \frac{u}{2}\right] \tag{A25}$$

The first term in (A25) contributes to the influx in the classical Hertz-Knudsen equation. The second term is related to the correction Schrage made, although he did not take this approach. The complementary error function was also seen in Carey's book (1), although I am not sure the original source. In fact, Schrage work seems to be equivalent to setting the integration limit for $v_z$ as [-∞, u], while we set the limit for $v_z$ as [-∞, 0] . Since $v_z$-u represents random velocity of molecules relative to the average flow while $v_z$ represents absolute random velocity, when $v_z<0$, molecules indeed flow towards interface. Hence, the correct integration limit is what we used and the extra error function terms in (A22) appear as a consequence.

Next, we consider the density gradient term:

$$\int_{-\infty}^\infty dv_x \int_{-\infty}^\infty dv_y \int_{-\infty}^0 f_d \tau v_z^2 \left\{\frac{1}{n}\frac{dn}{dz}\right\} dv_z$$

$$= \tau n\left\{\frac{1}{n}\frac{dn}{dz}\right\}\left[\frac{m}{2\pi k_B T(z)}\right]^{3/2}\int_o^\infty 2\pi v_{//} exp\left[-\frac{mv_{//}^2}{2k_B T}\right]dv_{//}\int_{-\infty}^0 v_z^2 exp\left[-\frac{m(v_z-u)^2}{2k_B T}\right]dv_z$$

$$= \tau n \left\{\frac{1}{n}\frac{dn}{dz}\right\}\left[\frac{m}{2\pi k_B T}\right]^{3/2} \pi\left(\frac{2k_BT}{m}\right)\int_{-\infty}^{0}\left[(v_z-u)^2 \ + 2u(v_z-u)+u^2\right] exp\left[-\frac{m(v_z-u)^2}{2k_BT}\right] dv_z$$

$$= \tau n \left\{\frac{1}{n}\frac{dn}{dz}\right\}\left[\frac{m}{2\pi k_B T}\right]^{3/2} \pi\left(\frac{2k_BT}{m}\right)\left[\int_{-\infty}^{-u} v_z'^2 exp\left(-\frac{mv_z'^2}{2k_BT}\right) dv'_z + 2u\int_{-\infty}^{-u} v_z' exp\left(-\frac{mv_z'^2}{2k_BT}\right) dv'_z + u^2\int_{-\infty}^{-u} exp\left(-\frac{mv_z'^2}{2k_BT}\right) dv'_z\right]$$

$$= \tau n \left\{\frac{1}{n}\frac{dn}{dz}\right\}\left[\frac{m}{2\pi k_B T}\right]^{3/2} \pi\left(\frac{2k_BT}{m}\right)\left[\int_{u}^{\infty} v_z'^2 exp\left(-\frac{mv_z'^2}{2k_BT}\right) dv'_z - 2u\int_{u}^{\infty} v_z' exp\left(-\frac{mv_z'^2}{2k_BT}\right) dv'_z + u^2\int_{u}^{\infty} exp\left(-\frac{mv_z'^2}{2k_BT}\right) dv'_z\right]$$

$$= \tau n \left\{\frac{1}{n}\frac{dn}{dz}\right\}\pi\left(\frac{2k_BT}{m}\right)\left[\frac{m}{2\pi k_B T}\right]^{3/2}\left[\frac{1}{2}\left(\frac{2k_BT}{m}\right)^{3/2}\left(\delta e^{-\delta^2}+\frac{\sqrt{\pi}}{2}erfc(\delta)\right) - u\left(\frac{2k_BT}{m}\right)e^{-\delta^2} + \frac{u^2}{2}\sqrt{\frac{2\pi k_BT}{m}}\, erfc(\delta)\right]$$

$$= \tau n \left\{\frac{1}{n}\frac{dn}{dz}\right\}\left(\frac{2k_BT}{\sqrt{\pi}\, m}\right)\left[\frac{1}{2}\left(\delta e^{-\delta^2}+\frac{\sqrt{\pi}}{2}erfc(\delta)\right) - \delta e^{-\delta^2} + \delta^2\frac{\sqrt{\pi}}{2}erfc(\delta)\right]$$

$$= \tau n \left\{\frac{1}{n}\frac{dn}{dz}\right\}\left(\frac{2k_BT}{\sqrt{\pi}\, m}\right)\left[-\frac{1}{2}\delta e^{-\delta^2}+\frac{\sqrt{\pi}}{2}\left(\frac{1}{2}+\delta^2\right)erfc(\delta)\right] \qquad \text{(A26)}$$

$$\approx \tau\frac{dn}{dz}\left(\frac{k_BT}{2m}\right) \qquad \text{(A27)}$$

The last approximation equals half of Eq. (A11).

Next, we consider dT/dz term

$$\int_{-\infty}^{\infty} dv_x \int_{-\infty}^{\infty} dv_y \int_{-\infty}^{0} f_d \tau v_z^2\left\{\frac{1}{T}\frac{dT}{dz}\left[\frac{m\left[v_x^2+v_y^2+(v_z-u(z))^2\ \right]}{2k_BT(z)}-\frac{3}{2}\right]\right\} dv_z$$

$$= n\tau\left[\frac{m}{2\pi k_BT}\right]^{3/2}\frac{1}{T}\frac{dT}{dz}\int_{-\infty}^{\infty} dv_x\int_{-\infty}^{\infty} dv_y\int_{-\infty}^{0} v_z^2\left\{\left[\frac{m\left[v_x^2+v_y^2+(v_z-u)^2\ \right]}{2k_BT}-\frac{3}{2}\right]\right\} dv_z$$

$$= n\tau\left[\frac{m}{2\pi k_BT}\right]^{3/2}\frac{1}{T}\frac{dT}{dz}\left[\frac{m}{2k_BT}\int_{o}^{\infty} 2\pi {v_{//}}^3 exp\left(-\frac{mv_{//}^2}{2k_BT}\right) dv_{//}\int_{-\infty}^{0} v_z^2 exp\left(-\frac{m(v_z-u)^2}{2k_BT}\right) dv_z\right.$$

$$+\frac{m}{2k_BT}\int_{o}^{\infty} 2\pi v_{//}\ \ exp\left(-\frac{mv_{//}^2}{2k_BT}\right) dv_{//}\int_{-\infty}^{0} v_z^2(v_z-u)^2\ exp\left(-\frac{m(v_z-u)^2}{2k_BT}\right) dv_z$$

$$\left.-\frac{3}{2}\int_{o}^{\infty} 2\pi v_{//}\ \ exp\left(-\frac{mv_{//}^2}{2k_BT}\right) dv_{//}\int_{-\infty}^{0} v_z^2 exp\left(-\frac{m(v_z-u)^2}{2k_BT}\right) dv_z\right]$$

$$= n\tau\left[\frac{m}{2\pi k_B T}\right]^{3/2}\frac{1}{T}\frac{dT}{dz}\left[\frac{m}{2k_B T}\pi\left(\frac{2k_B T}{m}\right)^2\int_{-\infty}^{-u}[v_z'^2+2uv_z'+u^2]exp\left(-\frac{mv_z'^2}{2k_B T}\right)dv'_z\right.$$

$$+\frac{m}{2k_B T}\pi\left(\frac{2k_B T}{m}\right)\int_{-\infty}^{-u}[v_z'^4+2uv_z'^3+u^2v_z'^2]exp\left(-\frac{mv_z'^2}{2k_B T}\right)dv'_z$$

$$\left.-\frac{3}{2}\pi\left(\frac{2k_B T}{m}\right)\int_{-\infty}^{-u}[v_z'^2+2uv_z'+u^2]exp\left(-\frac{mv_z'^2}{2k_B T}\right)dv'_z\right]$$

$$= n\tau\left[\frac{m}{2\pi k_B T}\right]^{3/2}\frac{1}{T}\frac{dT}{dz}\left[\frac{m}{2k_B T}\pi\left(\frac{2k_B T}{m}\right)^2\int_{u}^{\infty}[v_z'^2-2uv_z'+u^2]exp\left(-\frac{mv_z'^2}{2k_B T}\right)dv'_z\right.$$

$$+\frac{m}{2k_B T}\pi\left(\frac{2k_B T}{m}\right)\int_{u}^{\infty}[v_z'^4-2uv_z'^3+u^2v_z'^2]exp\left(-\frac{mv_z'^2}{2k_B T}\right)dv'_z$$

$$\left.-\frac{3}{2}\pi\left(\frac{2k_B T}{m}\right)\int_{u}^{\infty}[v_z'^2-2uv_z'+u^2]exp\left(-\frac{mv_z'^2}{2k_B T}\right)dv'_z\right]$$

$$= n\tau\left[\frac{m}{2\pi k_B T}\right]^{3/2}\frac{1}{T}\frac{dT}{dz}\left[\frac{m}{2k_B T}\pi\left(\frac{2k_B T}{m}\right)^2\left(\frac{2k_B T}{m}\right)^{3/2}\left(-\frac{1}{2}\delta e^{-\delta^2}+\frac{\sqrt{\pi}}{2}\left(\frac{1}{2}+\delta^2\right)erfc(\delta)\right)\right.$$

$$+\frac{m}{2k_B T}\pi\left(\frac{2k_B T}{m}\right)\left(\frac{2k_B T}{m}\right)^{5/2}\int_{\delta}^{\infty}[x^4-2\delta x^3+\delta^2x^2]exp(-x^2)dx$$

$$\left.-\frac{3}{2}\pi\left(\frac{2k_B T}{m}\right)\left(\frac{2k_B T}{m}\right)^{3/2}\int_{\delta}^{\infty}\left[x^2-2\delta x\quad+\delta^2\right]exp(-x^2)dx\right]$$

$$= n\tau\left[\frac{m}{2\pi k_B T}\right]^{3/2}\frac{1}{T}\frac{dT}{dz}\left[\frac{m}{2k_B T}\pi\left(\frac{2k_B T}{m}\right)^2\left(\frac{2k_B T}{m}\right)^{3/2}\langle-\frac{1}{2}\delta e^{-\delta^2}+\frac{\sqrt{\pi}}{2}\left(\frac{1}{2}+\delta^2\right)erfc(\delta)\rangle\right.$$

$$+\frac{m}{2k_B T}\pi\left(\frac{2k_B T}{m}\right)\left(\frac{2k_B T}{m}\right)^{5/2}\left[\left(\frac{\delta^3}{2}e^{-\delta^2}+\frac{3\delta}{4}e^{-\delta^2}+\frac{3\sqrt{\pi}}{8}erfc(\delta)\right)-\delta(1+\delta^2)e^{-\delta^2}+\right.$$

$$\left.\left.\delta^2\left(\frac{1}{2}\delta e^{-\delta^2}+\frac{\sqrt{\pi}}{4}erfc(\delta)\right)\right]-\frac{3}{2}\pi\left(\frac{2k_B T}{m}\right)\left(\frac{2k_B T}{m}\right)^{3/2}\left(-\frac{1}{2}\delta e^{-\delta^2}+\frac{\sqrt{\pi}}{2}\left(\frac{1}{2}+\delta^2\right)erfc(\delta)\right)\right]$$

$$= n\tau\frac{1}{\sqrt{\pi}}\frac{1}{T}\frac{dT}{dz}\left(\frac{2k_B T}{m}\right)\left[\left(-\frac{1}{2}\delta e^{-\delta^2}+\frac{\sqrt{\pi}}{2}\left(\frac{1}{2}+\delta^2\right)erfc(\delta)\right)\right.$$

$$+\left[\left(\frac{\delta^3}{2}e^{-\delta^2}+\frac{3\delta}{4}e^{-\delta^2}+\frac{3\sqrt{\pi}}{8}erfc(\delta)\right)-\delta(1+\delta^2)e^{-\delta^2}+\delta^2\left(\frac{1}{2}\delta e^{-\delta^2}+\frac{\sqrt{\pi}}{4}erfc(\delta)\right)\right]$$

$$\left.-\frac{3}{2}\left(-\frac{1}{2}\delta e^{-\delta^2}+\frac{\sqrt{\pi}}{2}\left(\frac{1}{2}+\delta^2\right)erfc(\delta)\right)\right]$$

$$\approx n\tau\left(\frac{k_BT}{2m}\right)\frac{1}{T}\frac{dT}{dz}erfc(\delta) \quad \text{(A28)}$$

$$\approx n\tau\frac{1}{T}\frac{dT}{dz}\left(\frac{k_BT}{2m}\right) \quad \text{(A29)}$$

Eq. (A29) is half of Eq. (A12), showing consistency. We will see next in the du/dz term.

$$\int_{-\infty}^{\infty} dv_x \int_{-\infty}^{\infty} dv_y \int_{-\infty}^{0} f_d \tau v_z^2 \left\{\frac{m[v_z-u(z)]}{k_BT(z)}\frac{du}{dz}\right\} dv_z$$

$$= \tau n\frac{m}{k_BT}\frac{du}{dz}\int_o^{\infty} 2\pi v_{//} \quad exp\left(-\frac{mv_{//}^2}{2k_BT}\right) dv_{//} \int_{-\infty}^{0}(v_z-u)v_z^2 exp\left(-\frac{m(v_z-u)^2}{2k_BT}\right) dv_z$$

$$= \tau n\frac{m}{k_BT}\left[\frac{m}{2\pi k_BT}\right]^{3/2}\frac{du}{dz}\int_o^{\infty} 2\pi v_{//} \quad exp\left(-\frac{mv_{//}^2}{2k_BT}\right) dv_{//} \int_{-\infty}^{0}(v_z-u)v_z^2 exp\left(-\frac{m(v_z-u)^2}{2k_BT}\right) dv_z$$

$$= \tau n\frac{m}{k_BT}\left[\frac{m}{2\pi k_BT}\right]^{3/2}\frac{du}{dz}\pi\left(\frac{2k_BT}{m}\right)\int_{-\infty}^{0}[(v_z-u)^3 + 2u(v_z-u)^2 + u^2(v_z - u)]exp\left(-\frac{m(v_z-u)^2}{2k_BT}\right) dv_z$$

$$= \tau n\frac{m}{k_BT}\left[\frac{m}{2\pi k_BT}\right]^{3/2}\frac{du}{dz}\pi\left(\frac{2k_BT}{m}\right)\int_{-\infty}^{-u}[v_z'^3 + 2uv_z'^2 + u^2v_z']exp\left(-\frac{mv_z'^2}{2k_BT}\right) dv_z'$$

$$= \tau n\frac{m}{k_BT}\left[\frac{m}{2\pi k_BT}\right]^{3/2}\frac{du}{dz}\pi\left(\frac{2k_BT}{m}\right)\left(\frac{2k_BT}{m}\right)^2\int_{\delta}^{\infty}[-x^3 + 2\delta x^2 - \delta^2 x]exp(-x^2)dx$$

$$= \tau n\frac{2}{\sqrt{\pi}}\frac{du}{dz}\left(\frac{2k_BT}{m}\right)^{1/2}\left[-\frac{1}{2}(1+\delta^2)e^{-\delta^2} + 2\delta\left(\frac{1}{2}\delta e^{-\delta^2} + \frac{\sqrt{\pi}}{4}erfc(\delta)\right) - \frac{1}{2}\delta^2 e^{-\delta^2}\right]$$

$$= \tau n\frac{2}{\sqrt{\pi}}\frac{du}{dz}\left(\frac{2k_BT}{m}\right)^{1/2}\left[-\frac{1}{2}e^{-\delta^2} + \frac{\sqrt{\pi}}{2}\delta erfc(\delta)\right] \quad \text{(A30)}$$

$$\approx -\tau n\frac{1}{\sqrt{\pi}}\frac{du}{dz}\left(\frac{2k_BT}{m}\right)^{1/2} + \tau nu\frac{du}{dz} \quad \text{(A31)}$$

It is interesting that the first term is not in Eq. (A13) because the odd velocity terms do not cancel out. The coefficient for this term much larger. If this term is important, likely we need higher order terms in Eq. (A14) since the second term is exactly half of the corresponding term in Eq.(A14).

Summing up all of the above terms, and applying the approximations used, the flux going towards the interface is then

$$J_m^- = -n\left[\left(\frac{k_BT}{2\pi m}\right)^{1/2} - \frac{u}{2}\right] - \tau\frac{dn}{dz}\left(\frac{k_BT}{2m}\right) - n\tau\frac{1}{T}\frac{dT}{dz}\left(\frac{k_BT}{2m}\right) + \tau n\frac{1}{\sqrt{\pi}}\frac{du}{dz}\left(\frac{2k_BT}{m}\right)^{1/2} - \tau nu\frac{du}{dz}$$

$$\approx -n\left[\left(\frac{k_BT}{2\pi m}\right)^{1/2}-\frac{u}{2}\right]-\tau\frac{dn}{dz}\left(\frac{k_BT}{2m}\right)-n\tau\frac{1}{T}\frac{dT}{dz}\left(\frac{k_BT}{2m}\right) \quad \text{(A32)}$$

$$= -n\left(\frac{k_BT}{2\pi m}\right)^{1/2}+\frac{J_m}{2} \quad \text{(A33)}$$

Where in Eq. (A32) step, we assume that du/dz terms are negligible, which is reasonable for one-dimensional problems but may not be a good approximation for multidimensional problems since first term in Eq. (31) is much larger than the second term if du/dz is not negligible. In writing down Eq. (A33), we used Eq. (A15).

**Heat Flux Coming Towards Interface.** The heat flux has convection term, the density, temperature, and velocity gradient terms, which are evaluated as below.

Convection term

$$\int_{-\infty}^{\infty} dv_x \int_{-\infty}^{\infty} dv_y \int_{-\infty}^{0} v_z \frac{m}{2}\left[v_x^2+v_y^2+v_z^2\right] f_d dv_z$$

$$= \frac{m}{2} n\left[\frac{m}{2\pi k_BT}\right]^{3/2}\left[\int_o^{\infty} 2\pi v_{//}{}^3 exp\left(-\frac{mv_{//}^2}{2k_BT}\right) dv_{//}\int_{-\infty}^{0} v_z exp\left(-\frac{m(v_z-u)^2}{2k_BT}\right) dv_z\right.$$

$$\left. + \int_o^{\infty} 2\pi v_{//} \;\; exp\left(-\frac{mv_{//}^2}{2k_BT}\right) dv_{//}\int_{-\infty}^{0} v_z{}^3 exp\left(-\frac{m(v_z-u)^2}{2k_BT}\right) dv_z\right]$$

$$= \frac{m}{2} n\left[\frac{m}{2\pi k_BT}\right]^{3/2}\left[\pi\left(\frac{2k_BT}{m}\right)^2 \int_{-\infty}^{-u}[v'_z+u] exp\left(-\frac{mv_z'^2}{2k_BT}\right) dv_z\right.$$

$$\left. +\pi\left(\frac{2k_BT}{m}\right) \;\; \int_{-\infty}^{-u}[v_z{}'^3+3uv_z{}'^2+3u^2v_z{}'+u^3] exp\left(-\frac{mv_z'^2}{2k_BT}\right) dv_z\right]$$

$$= \frac{m}{2} n\left[\frac{m}{2\pi k_BT}\right]^{3/2}\left[\pi\left(\frac{2k_BT}{m}\right)^3 \int_{\delta}^{\infty}[-x+\delta] exp(-x^2) dx\right.$$

$$\left. +\pi\left(\frac{2k_BT}{m}\right)^3 \int_{\delta}^{\infty}[-x^3+3\delta x^2-3\delta^2 x+\delta^3] exp(-x^2) dx\right]$$

$$= \frac{m}{2\sqrt{\pi}} n\left(\frac{2k_BT}{m}\right)^{3/2}\left[\left(-\frac{1}{2}e^{-\delta^2}+\frac{\sqrt{\pi}}{2}\delta erfc(\delta)\right)+\left[-\frac{1}{2}(1+\delta^2)e^{-\delta^2}+\right.\right.$$

$$\left.\left.\left(\frac{3}{2}\delta^2 e^{-\delta^2}+\frac{3\sqrt{\pi}}{4}\delta erfc(\delta)\right)-\frac{3}{2}\delta^2 e^{-\delta^2}+\frac{\sqrt{\pi}}{2}\delta^3 erfc(\delta)\right]\right]$$

$$= \frac{m}{2\sqrt{\pi}} n\left(\frac{2k_BT}{m}\right)^{3/2}\left[-\left(1+\frac{\delta^2}{2}\right)e^{-\delta^2}+\frac{5\sqrt{\pi}}{4}\delta erfc(\delta)+\frac{\sqrt{\pi}}{2}\delta^3 erfc(\delta)\right] \quad \text{(A34)}$$

$$\approx -\frac{m}{2\sqrt{\pi}} n \left(\frac{2k_BT}{m}\right)^{3/2} + \frac{5mn}{8}\left(\frac{2k_BT}{m}\right) \quad u \tag{A35}$$

The second term in (A35) is half of the value in Eq. (A17) (neglecting the higher order term). An extra first term appears, which will be responsible for the temperature discontinuity.

dn/dz term is as follows

$$\int_{-\infty}^{\infty} dv_x \int_{-\infty}^{\infty} dv_y \int_{-\infty}^{0} f_d \tau v_z^2 \frac{m}{2}\left[v_x^2 + v_y^2 + v_z^2\right]\left\{\frac{1}{n}\frac{dn}{dz}\right\} dv_z$$

$$= \frac{m}{2} n\tau \left[\frac{m}{2\pi k_BT}\right]^{3/2} \frac{1}{n}\frac{dn}{dz}\left[\int_o^{\infty} 2\pi v_{//}{}^3 exp\left(-\frac{mv_{//}^2}{2k_BT}\right) dv_{//} \int_{-\infty}^{0} v_z{}^2 exp\left(-\frac{m(v_z-u)^2}{2k_BT}\right) dv_z\right.$$

$$\left. + \int_o^{\infty} 2\pi v_{//} \quad exp\left(-\frac{mv_{//}^2}{2k_BT}\right) dv_{//} \int_{-\infty}^{0} v_z{}^4 exp\left(-\frac{m(v_z-u)^2}{2k_BT}\right) dv_z\right]$$

$$= \frac{m}{2} n\tau \left[\frac{m}{2\pi k_BT}\right]^{3/2} \frac{1}{n}\frac{dn}{dz}\left[\pi\left(\frac{2k_BT}{m}\right)^2 \int_{-\infty}^{-u}[v_z{}'^2 + 2uv_z{}' + u^2] exp\left(-\frac{mv_z{}'^2}{2k_BT}\right) dv_z{}'\right.$$

$$\left. + \pi\left(\frac{2k_BT}{m}\right) \quad \int_{-\infty}^{-u}[v_z{}'^4 + 2u^2 v_z'^2 + u^4] exp\left(-\frac{mv_z{}'^2}{2k_BT}\right) dv_z{}'\right]$$

$$= \frac{m}{2} n\tau \left[\frac{m}{2\pi k_BT}\right]^{3/2} \frac{1}{n}\frac{dn}{dz}\pi\left(\frac{2k_BT}{m}\right)^{7/2} \left[\int_{\delta}^{\infty}[x^2 - 2\delta x + \delta^2] exp(-x^2) dx\right.$$

$$\left. + \int_{\delta}^{\infty}[x^4 + 2\delta^2 x^2 + \delta^4] exp(-x^2) dx\right]$$

$$= \frac{m}{2\sqrt{\pi}} n\tau \frac{1}{n}\frac{dn}{dz}\left(\frac{2k_BT}{m}\right)^2 \left[\left(-\frac{1}{2}\delta e^{-\delta^2} + \frac{\sqrt{\pi}}{2}\left(\frac{1}{2} + \delta^2\right) erfc(\delta)\right)\right.$$

$$\left. \left(\frac{\delta^3}{2} e^{-\delta^2} + \frac{3\delta}{4} e^{-\delta^2} + \frac{3\sqrt{\pi}}{8} erfc(\delta)\right) + \delta^2\left(\delta e^{-\delta^2} + \frac{\sqrt{\pi}}{2} erfc(\delta)\right) + \delta^4 \frac{\sqrt{\pi}}{2} erfc(\delta)\right]$$

$$= \frac{m}{2\sqrt{\pi}} n\tau \frac{1}{n}\frac{dn}{dz}\left(\frac{2k_BT}{m}\right)^2 \left[\frac{\delta}{4} e^{-\delta^2} + \frac{3\delta^3}{2} e^{-\delta^2} + \frac{\sqrt{\pi}}{2}\left(\frac{5}{4} + 2\delta^2 + \delta^4\right) erfc(\delta)\right] \tag{A36}$$

$$\approx \frac{m}{8\sqrt{\pi}} \tau \left(\frac{2k_BT}{m}\right)^{3/2} u \frac{dn}{dz} + \frac{5k_B{}^2T^2}{4m} \tau \frac{dn}{dz} \tag{A37}$$

$$\approx \frac{5k_B{}^2T^2}{4m} \tau \frac{dn}{dz} \tag{A38}$$

The first term in Eq.(A37) should be small since u is much smaller than thermal velocity and can be neglected, leading to Eq.(A38), which is half of the first term in Eq.(A18).

The temperature gradient term is:

$$\int_{-\infty}^{\infty} dv_x \int_{-\infty}^{\infty} dv_y \int_{-\infty}^{0} f_a \tau v_z^2 \frac{m}{2}\left[v_x^2+v_y^2+v_z^2\right]\frac{1}{T}\frac{dT}{dz}\left[\frac{m\left[v_x^2+v_y^2+(v_z-u(z))^2\right]}{2k_BT(z)}-\frac{3}{2}\right]dv_z$$

$$=\frac{m}{2}n\tau\left[\frac{m}{2\pi k_BT}\right]^{3/2}\frac{1}{T}\frac{dT}{dz}\left[\int_o^{\infty}2\pi v_{//}exp\left(-\frac{mv_{//}^2}{2k_BT}\right)dv_{//}\int_{-\infty}^{0}v_z^2\left[v_{//}^2+v_z^2\right]\left[\frac{m\left[v_{//}^2+(v_z-u)^2\right]}{2k_BT}-\frac{3}{2}\right]exp\left(-\frac{m(v_z-u)^2}{2k_BT}\right)dv_z\right.$$

$$=\frac{m}{2}n\tau\left[\frac{m}{2\pi k_BT}\right]^{3/2}\frac{1}{T}\frac{dT}{dz}\times\left[\frac{m}{2k_BT}\langle\int_o^{\infty}2\pi {v_{//}}^5 exp\left(-\frac{mv_{//}^2}{2k_BT}\right)dv_{//}\int_{-\infty}^{0}v_z^2 exp\left(-\frac{m(v_z-u)^2}{2k_BT}\right)dv_z\rangle\right.$$

$$+\frac{m}{2k_BT}\langle\int_o^{\infty}2\pi {v_{//}}^3 exp\left(-\frac{mv_{//}^2}{2k_BT}\right)dv_{//}\int_{-\infty}^{0}\left[v_z^4+v_z^2(v_z-u)^2\right]exp\left(-\frac{m(v_z-u)^2}{2k_BT}\right)dv_z\rangle$$

$$+\frac{m}{2k_BT}\langle\int_o^{\infty}2\pi {v_{//}}^1 exp\left(-\frac{mv_{//}^2}{2k_BT}\right)dv_{//}\int_{-\infty}^{0}v_z^4(v_z-u)^2\ exp\left(-\frac{m(v_z-u)^2}{2k_BT}\right)dv_z\rangle$$

$$-\frac{3}{2}\langle\int_o^{\infty}2\pi {v_{//}}^3 exp\left(-\frac{mv_{//}^2}{2k_BT}\right)dv_{//}\int_{-\infty}^{0}v_z^2 exp\left(-\frac{m(v_z-u)^2}{2k_BT}\right)dv_z\rangle$$

$$\left.-\frac{3}{2}\int_o^{\infty}2\pi v_{//}\ \ exp\left(-\frac{mv_{//}^2}{2k_BT}\right)dv_{//}\int_{-\infty}^{0}v_z^4 exp\left(-\frac{m(v_z-u)^2}{2k_BT}\right)dv_z\right]$$

$$=\frac{m}{2}n\tau\left[\frac{m}{2\pi k_BT}\right]^{3/2}\frac{1}{T}\frac{dT}{dz}\times\left[\frac{m}{2k_BT}2\pi\left(\frac{2k_BT}{m}\right)^3\langle\int_{-\infty}^{-u}\left[v_z'^2+2uv_z+u^2\right]exp\left(-\frac{m{v_z}'^2}{2k_BT}\right)dv_z\rangle\right.$$

$$+\frac{m}{2k_BT}\pi\left(\frac{2k_BT}{m}\right)^2\langle\int_{-\infty}^{-u}\left[2v_z'^4+3u^2v_z'^2+2uv_z'^3+u^4\right]exp\left(-\frac{m{v_z}'^2}{2k_BT}\right)dv'_z\rangle$$

$$+\frac{m}{2k_BT}\pi\left(\frac{2k_BT}{m}\right)\ \ \langle\int_{-\infty}^{-u}\left[v_z'^6+2u^2v_z'^4+u^4v_z'^2\right]exp\left(-\frac{m{v_z}'^2}{2k_BT}\right)dv_z\rangle$$

$$-\frac{3}{2}\pi\left(\frac{2k_BT}{m}\right)^2\langle\int_{-\infty}^{-u}\left[v_z'^2+2uv'_z+u^2\right]exp\left(-\frac{m{v_z}'^2}{2k_BT}\right)dv_z\rangle$$

$$\left.-\frac{3}{2}\pi\left(\frac{2k_BT}{m}\right)\ \ \int_{-\infty}^{-u}\left[v_z'^4+2u^4\ {v_z}'^2+u^4\right]exp\left(-\frac{m(v_z-u)^2}{2k_BT}\right)dv_z\right]$$

$$= \frac{m}{2} n\tau \left[\frac{m}{2\pi k_B T}\right]^{3/2} \frac{1}{T}\frac{dT}{dz} \times \left[\frac{m}{2k_B T} 2\pi \left(\frac{2k_B T}{m}\right)^{9/2} \langle \int_\delta^\infty [x^2 - 2\delta x + \delta^2] exp(-x^2) dx \rangle\right.$$

$$+ \frac{m}{2k_B T} \pi \left(\frac{2k_B T}{m}\right)^{9/2} \langle \int_\delta^\infty [2x^4 + 3\delta^2 x^2 - 2\delta x^3 + \delta^4] exp(-x^2) dx \rangle$$

$$+ \frac{m}{2k_B T} \pi \left(\frac{2k_B T}{m}\right)^{9/2} \langle \int_\delta^\infty [x^6 + 2\delta^2 x^4 + \delta^4 x^2] exp(-x^2) dx \rangle$$

$$- \frac{3}{2} \pi \left(\frac{2k_B T}{m}\right)^{7/2} \langle \int_\delta^\infty [x^2 - 2\delta x + \delta^2] exp(-x^2) dx \rangle$$

$$\left. - \frac{3}{2} \pi \left(\frac{2k_B T}{m}\right)^{7/2} \int_\delta^\infty [x^4 + 2\delta^2 x^2 + \delta^4] exp(-x^2) dx \right]$$

$$= \frac{m}{2\sqrt{\pi}} n\tau \left(\frac{2k_B T}{m}\right)^2 \frac{1}{T}\frac{dT}{dz} \times \left[2\left(-\frac{1}{2}\delta e^{-\delta^2} + \frac{\sqrt{\pi}}{2}\left(\frac{1}{2} + \delta^2\right) erfc(\delta)\right)\right.$$

$$+ \left[2\left(\frac{\delta^3}{2} e^{-\delta^2} + \frac{3\delta}{4} e^{-\delta^2} + \frac{3\sqrt{\pi}}{8} erfc(\delta)\right) + \frac{3\delta^2}{2}\left(\delta e^{-\delta^2} + \frac{\sqrt{\pi}}{2} erfc(\delta)\right) - \delta(1 + \delta^2) e^{-\delta^2} + \delta^4 \frac{\sqrt{\pi}}{2} erfc(\delta)\right]$$

$$+ \left[\left(\frac{\delta^5}{2} e^{-\delta^2} + \frac{5\delta^3}{4} e^{-\delta^2} + \frac{15\delta}{8} e^{-\delta^2} + \frac{15\sqrt{\pi}}{16} erfc(\delta)\right) + 2\delta^2 \left(\frac{\delta^3}{2} e^{-\delta^2} + \frac{3\delta}{4} e^{-\delta^2} + \frac{3\sqrt{\pi}}{8} erfc(\delta)\right) + \delta^4 \left(\frac{\delta}{2} e^{-\delta^2} + \frac{\sqrt{\pi}}{4} erfc(\delta)\right)\right]$$

$$- \frac{3}{2}\left(-\frac{1}{2}\delta e^{-\delta^2} + \frac{\sqrt{\pi}}{2}\left(\frac{1}{2} + \delta^2\right) erfc(\delta)\right)$$

$$\left. - \frac{3}{2}\left[\left(\frac{\delta^3}{2} e^{-\delta^2} + \frac{3\delta}{4} e^{-\delta^2} + \frac{3\sqrt{\pi}}{8} erfc(\delta)\right) + \delta^2 \left(\delta e^{-\delta^2} + \frac{\sqrt{\pi}}{2} erfc(\delta)\right) + \delta^4 \frac{\sqrt{\pi}}{2} erfc(\delta)\right]\right]$$

$$= \frac{m}{2\sqrt{\pi}} n\tau \left(\frac{2k_B T}{m}\right)^2 \frac{1}{T}\frac{dT}{dz} \left[(\delta + 2\delta^3 + 2\delta^5) e^{-\delta^2} + \sqrt{\pi}\left(\frac{5}{4} + \delta^2\right) erfc(\delta)\right] \qquad \text{(A39)}$$

$$\approx \frac{m}{2\sqrt{\pi}} n\tau \left(\frac{2k_B T}{m}\right)^{3/2} \frac{u}{T}\frac{dT}{dz} + \frac{5m}{8} n\tau \left(\frac{2k_B T}{m}\right)^2 \frac{1}{T}\frac{dT}{dz} \qquad \text{(A40)}$$

Eq.(A40) is half of Eq.(A19).

The final term, the du/dz term is

$$\int_{-\infty}^{\infty} dv_x \int_{-\infty}^{\infty} dv_y \int_{-\infty}^{0} f_d \tau v_z^2 \frac{m}{2}\left[v_x^2 + v_y^2 + v_z^2\right] \frac{m[v_z - u(z)]}{k_B T(z)} \frac{du}{dz} dv_z$$

$$= \frac{m}{2}\tau n \frac{m}{k_B T}\frac{du}{dz}\left[\frac{m}{2\pi k_B T}\right]^{3/2} \left[\int_o^{\infty} 2\pi {v_{//}}^3 exp\left(-\frac{mv_{//}^2}{2k_B T}\right) dv_{//} \int_{-\infty}^{0} v_z^2 [v_z - u(z)] exp\left(-\frac{m(v_z-u)^2}{2k_B T}\right) dv_z\right.$$

$$\left. + \int_o^{\infty} 2\pi v_{//} \quad exp\left(-\frac{mv_{//}^2}{2k_B T}\right) dv_{//} \int_{-\infty}^{0} v_z^4 [v_z - u(z)] exp\left(-\frac{m(v_z-u)^2}{2k_B T}\right) dv_z\right]$$

$$= \frac{m}{2}\tau n \frac{m}{k_B T}\frac{du}{dz}\left[\frac{m}{2\pi k_B T}\right]^{3/2} \left[\pi\left(\frac{2k_B T}{m}\right)^4 \int_{\delta}^{\infty} \quad [-x^3 + 2\delta x^2 - \delta^2 x] exp(-x^2) dx\right.$$

$$\left. + \pi\left(\frac{2k_B T}{m}\right)^4 \int_{\delta}^{\infty} [-x^5 - 2\delta^2 x^3 - \delta^4 x] exp(-x^2) dx\right]$$

$$= \frac{m}{\sqrt{\pi}}\tau n \frac{du}{dz}\left(\frac{2k_B T}{m}\right)^{3/2} \left[ \quad \int_{\delta}^{\infty} [-x^3 + 2\delta x^2 - \delta^2 x] exp(-x^2)\, dx\right.$$

$$\left. + \int_{\delta}^{\infty} [-x^5 - 2\delta^2 x^3 - \delta^4 x] exp(-x^2) dx\right]$$

$$= \frac{m}{\sqrt{\pi}}\tau n \frac{du}{dz}\left(\frac{2k_B T}{m}\right)^{3/2} \left[\left(-\delta^2 e^{-\delta^2} + \frac{\sqrt{\pi}}{2}\delta erfc(\delta)\right) - \left[2\delta^4 + 3\delta^2 + \frac{3}{2}\right] e^{-\delta^2}\right]$$

$$\approx \frac{m}{\sqrt{\pi}}\tau n \frac{du}{dz}\left(\frac{2k_B T}{m}\right)^{3/2} \left(-\frac{3}{2} + \frac{\sqrt{\pi}}{2}\delta\right)$$

$$\approx -\frac{3m}{2\sqrt{\pi}}\tau n \frac{du}{dz}\left(\frac{2k_B T}{m}\right)^{3/2} + \tau n u k_B T \frac{du}{dz} \qquad \text{(A42)}$$

The second term is half of Eq. (A20). However, the first term can be much larger than the second term. In 1D situation, we expect that du/dz is small and hence both terms can be neglected.

Combining all energy flux terms, the heat flux coming towards interface is

$$q^- = -\frac{m}{2\sqrt{\pi}} n \left(\frac{2k_B T}{m}\right)^{3/2} + \frac{5mn}{8}\left(\frac{2k_B T}{m}\right) \quad u - \frac{5{k_B}^2 T^2}{4m}\tau \frac{dn}{dz} - n\tau \frac{1}{T}\frac{dT}{dz}\left(\frac{k_B T}{2m}\right) \qquad \text{(A43)}$$

$$\approx -\frac{m}{2\sqrt{\pi}} n \left(\frac{2k_B T}{m}\right)^{3/2} + \frac{q}{2} \qquad \text{(A44)}$$

**Heat Flux Emitted from Interface to Vapor Phase**:

$$q^+ = \alpha \int_{-\infty}^{\infty} dv_x \int_{-\infty}^{\infty} dv_y \int_0^{\infty} v_z \frac{mv^2}{2} f_e dv_z$$

$$= \frac{m}{2}\alpha n_s(T_l)\left[\frac{m}{2\pi k_B T_l}\right]^{3/2}\int_{-\infty}^{\infty} dv_x \int_{-\infty}^{\infty} dv_y \int_0^{\infty} v_z\left[v_x^2 + v_y^2 + v_z^2\right] exp\left(-\frac{m\left[v_x^2+v_y^2+v_z^2\right]}{2k_B T_l}\right) dv_z$$

$$= \frac{m}{2}\alpha n_s(T_l)\left[\frac{m}{2\pi k_B T_l}\right]^{3/2}\pi\left(\frac{2k_B T}{m}\right)^2\left[\pi\left(\frac{2k_B T}{m}\right)^2\frac{1}{2}\left(\frac{2k_B T}{m}\right) \quad + \pi\left(\frac{2k_B T}{m}\right) \quad \frac{1}{2}\left(\frac{2k_B T}{m}\right)^2\right]$$

$$= \frac{m}{2\sqrt{\pi}}\alpha n_s(T_l)\left(\frac{2k_B T}{m}\right)^{3/2} \tag{A45}$$

**Density Contributed by Molecules Moving Towards Interface.** Now, we consider contributions of molecules moving towards the interface to the density at the interface.

$$n^-(z=0) = \int_{-\infty}^{\infty} dv_x \int_{-\infty}^{\infty} dv_y \int_{-\infty}^{o} f_d dv_z$$

$$- \int_{-\infty}^{\infty} dv_x \int_{-\infty}^{\infty} dv_y \int_{-\infty}^{0} f_d \tau v_z \left\{\frac{1}{n}\frac{dn}{dz} + \frac{1}{T}\frac{dT}{dz}\left[\frac{m\left[v_x^2+v_y^2+(v_z-u(z))^2\right]}{2k_B T(z)} - \frac{3}{2}\right] + \frac{m[v_z-u(z)]}{k_B T(z)}\frac{du}{dz}\right\} dv_z \tag{A46}$$

The first the term related to $f_d$ is

$$\int_{-\infty}^{\infty} dv_x \int_{-\infty}^{\infty} dv_y \int_{-\infty}^{0} f_d dv_z = \frac{n_v(0)}{2} \tag{A47}$$

Next, we consider the density gradient term:

$$\int_{-\infty}^{\infty} dv_x \int_{-\infty}^{\infty} dv_y \int_{-\infty}^{0} f_d \tau v_z \left\{\frac{1}{n}\frac{dn}{dz}\right\} dv_z$$

$$= \tau n\left\{\frac{1}{n}\frac{dn}{dz}\right\}\left[\frac{m}{2\pi k_B T(z)}\right]^{3/2}\int_o^{\infty} 2\pi v_{//} exp\left[-\frac{m v_{//}^2}{2k_B T}\right] dv_{//}\int_{-\infty}^{0} v_z \ exp\left[-\frac{m(v_z-u)^2}{2k_B T}\right] dv_z$$

$$= \tau n\left\{\frac{1}{n}\frac{dn}{dz}\right\}\left[\frac{m}{2\pi k_B T}\right]^{3/2}\pi\left(\frac{2k_B T}{m}\right)\int_{-\infty}^{0}\left[(v_z-u)+u\right] exp\left[-\frac{m(v_z-u)^2}{2k_B T}\right] dv_z$$

$$= \tau n\left\{\frac{1}{n}\frac{dn}{dz}\right\}\left[\frac{m}{2\pi k_B T}\right]^{3/2}\pi\left(\frac{2k_B T}{m}\right)\left[\int_{-\infty}^{-u} v_z' exp\left(-\frac{m v_z'^2}{2k_B T}\right) dv'_z + u\int_{-\infty}^{-u} exp\left(-\frac{m v_z'^2}{2k_B T}\right) dv'_z\right]$$

$$= \tau n\left\{\frac{1}{n}\frac{dn}{dz}\right\}\left[\frac{m}{2\pi k_B T}\right]^{3/2}\pi\left(\frac{2k_B T}{m}\right)\left[-\int_{u}^{\infty} v_z' exp\left(-\frac{m v_z'^2}{2k_B T}\right) dv'_z + u\int_{u}^{\infty} exp\left(-\frac{m v_z'^2}{2k_B T}\right) dv'_z\right]$$

$$= \tau n\left\{\frac{1}{n}\frac{dn}{dz}\right\}\pi\left(\frac{2k_B T}{m}\right)\left[\frac{m}{2\pi k_B T}\right]^{3/2}\left[-\frac{1}{2}\left(\frac{2k_B T}{m}\right)e^{-\delta^2} + \frac{u}{2}\sqrt{\frac{2\pi k_B T}{m}}\, erfc(\delta)\right]$$

$$= \tau n\left\{\frac{1}{n}\frac{dn}{dz}\right\}\left(\frac{2k_B T}{\pi m}\right)^{1/2}\left[-\frac{1}{2}e^{-\delta^2} + \frac{\sqrt{\pi}}{2}\delta erfc(\delta)\right]$$

$$\approx -\frac{1}{2}\tau\frac{dn}{dz}\left(\frac{2k_BT}{\pi\, m}\right)^{1/2} \tag{A48}$$

Next, we consider dT/dz term

$$\int_{-\infty}^{\infty}dv_x\int_{-\infty}^{\infty}dv_y\int_{-\infty}^{0}f_d\tau v_z\ \left\{\frac{1}{T}\frac{dT}{dz}\left[\frac{m\left[v_x^2+v_y^2+(v_z-u(z))^2\ \right]}{2k_BT(z)}-\frac{3}{2}\right]\right\}dv_z$$

$$=n\tau\left[\frac{m}{2\pi k_BT}\right]^{3/2}\frac{1}{T}\frac{dT}{dz}\int_{-\infty}^{\infty}dv_x\int_{-\infty}^{\infty}dv_y\int_{-\infty}^{0}v_z\ \left[\frac{m\left[v_x^2+v_y^2+(v_z-u)^2\ \right]}{2k_BT}-\frac{3}{2}\right]dv_z$$

$$=n\tau\left[\frac{m}{2\pi k_BT}\right]^{3/2}\frac{1}{T}\frac{dT}{dz}\left[\frac{m}{2k_BT}\int_o^{\infty}2\pi {v_{//}}^3exp\left(-\frac{mv_{//}^2}{2k_BT}\right)dv_{//}\int_{-\infty}^{0}v_z\ exp\left(-\frac{m(v_z-u)^2}{2k_BT}\right)dv_z\right.$$

$$+\frac{m}{2k_BT}\int_o^{\infty}2\pi v_{//}\ \ exp\left(-\frac{mv_{//}^2}{2k_BT}\right)dv_{//}\int_{-\infty}^{0}v_z\ (v_z-u)^2\ exp\left(-\frac{m(v_z-u)^2}{2k_BT}\right)dv_z$$

$$\left.-\frac{3}{2}\int_o^{\infty}2\pi v_{//}\ \ exp\left(-\frac{mv_{//}^2}{2k_BT}\right)dv_{//}\int_{-\infty}^{0}v_z\ exp\left(-\frac{m(v_z-u)^2}{2k_BT}\right)dv_z\right]$$

$$=n\tau\left[\frac{m}{2\pi k_BT}\right]^{3/2}\frac{1}{T}\frac{dT}{dz}\left[\frac{m}{2k_BT}\pi\left(\frac{2k_BT}{m}\right)^2\int_{-\infty}^{-u}[v_z'+u]exp\left(-\frac{mv_z'^2}{2k_BT}\right)dv'_z\right.$$

$$+\frac{m}{2k_BT}\pi\left(\frac{2k_BT}{m}\right)\int_{-\infty}^{-u}[v_z'^3+uv_z'^2]exp\left(-\frac{mv_z'^2}{2k_BT}\right)dv'_z$$

$$\left.-\frac{3}{2}\pi\left(\frac{2k_BT}{m}\right)\int_{-\infty}^{-u}[v_z'+u]exp\left(-\frac{mv_z'^2}{2k_BT}\right)dv'_z\right]$$

$$=n\tau\left[\frac{m}{2\pi k_BT}\right]^{3/2}\frac{1}{T}\frac{dT}{dz}\left[-\frac{\pi}{2}\left(\frac{2k_BT}{m}\right)\quad\int_u^{\infty}[v_z'-u]exp\left(-\frac{mv_z'^2}{2k_BT}\right)dv'_z\right.$$

$$\left.-\pi\int_u^{\infty}[v_z'^3-uv_z'^2]exp\left(-\frac{mv_z'^2}{2k_BT}\right)dv'_z\right]$$

$$=n\tau\left[\frac{m}{2\pi k_BT}\right]^{3/2}\frac{1}{T}\frac{dT}{dz}\left[-\frac{\pi}{2}\left(\frac{2k_BT}{m}\right)^2\left(\frac{1}{2}e^{-\delta^2}-\frac{\sqrt{\pi}}{2}\delta\ erfc(\delta)\right)\right.$$

$$\left.-\pi\left(\frac{2k_BT}{m}\right)^2\int_{\delta}^{\infty}[x^3-\delta x^2]exp(-x^2)dx\right]$$

$$=n\tau\frac{1}{T}\frac{dT}{dz}\left(\frac{2k_BT}{\pi m}\right)^{1/2}\left[-\frac{1}{2}\left(\frac{1}{2}e^{-\delta^2}-\frac{\sqrt{\pi}}{2}\delta\ erfc(\delta)\right)\right.$$

$$\left.-\left[\frac{1}{2}(1+\delta^2)e^{-\delta^2}-\delta\left(\frac{1}{2}\delta e^{-\delta^2}+\frac{\sqrt{\pi}}{4}erfc(\delta)\right)\right]\right]$$

$$\approx -\frac{3}{4} n\tau \frac{1}{T}\frac{dT}{dz}\left(\frac{2k_BT}{\pi m}\right)^{1/2} \quad \text{(A49)}$$

Next, we evaluate the du/dz term

$$\int_{-\infty}^{\infty} dv_x \int_{-\infty}^{\infty} dv_y \int_{-\infty}^{0} f_d \tau v_z \left\{\frac{m[v_z-u(z)]}{k_BT(z)}\frac{du}{dz}\right\} dv_z$$

$$= \tau n \frac{m}{k_BT}\left[\frac{m}{2\pi k_BT}\right]^{3/2} \frac{du}{dz}\int_o^{\infty} 2\pi v_{//} \ \ exp\left(-\frac{mv_{//}^2}{2k_BT}\right) dv_{//} \int_{-\infty}^{0} (v_z - u)v_z \ exp\left(-\frac{m(v_z-u)^2}{2k_BT}\right) dv_z$$

$$= \tau n \frac{m}{k_BT}\left[\frac{m}{2\pi k_BT}\right]^{3/2} \frac{du}{dz}\pi\left(\frac{2k_BT}{m}\right) \int_{-\infty}^{0} [(v_z-u)^2 + u(v_z-u)] exp\left(-\frac{m(v_z-u)^2}{2k_BT}\right) dv_z$$

$$= \tau n \frac{m}{k_BT}\left[\frac{m}{2\pi k_BT}\right]^{3/2} \frac{du}{dz}\pi\left(\frac{2k_BT}{m}\right) \int_{-\infty}^{-u} [v_z'^2 + uv_z'] exp\left(-\frac{mv_z'^2}{2k_BT}\right) dv_z'$$

$$= -\tau n \frac{m}{k_BT}\left[\frac{m}{2\pi k_BT}\right]^{3/2} \frac{du}{dz}\pi\left(\frac{2k_BT}{m}\right)\left(\frac{2k_BT}{m}\right)^{3/2} \int_{\delta}^{\infty} [-x^2 + \delta x] exp(-x^2) dx$$

$$= -\tau n \frac{2}{\sqrt{\pi}}\frac{du}{dz}\left[-\left(\frac{1}{2}\delta e^{-\delta^2} + \frac{\sqrt{\pi}}{4} erfc(\delta)\right) + \frac{1}{2}\delta e^{-\delta^2}\right]$$

$$\approx 0 \quad \text{(A50)}$$

So, the contribution of molecules in negative direction to the local density is

$$n^- = \frac{n_v(0)}{2} + \frac{1}{2}\tau\frac{dn}{dz}\left(\frac{2k_BT}{\pi m}\right)^{1/2} + \frac{3}{4} n\tau \frac{1}{T}\frac{dT}{dz}\left(\frac{2k_BT}{\pi m}\right)^{1/2} \quad \text{(A51)}$$

The forward going density consists of molecules emitted by the surface and reflected

$$n^+ = \alpha \frac{n_s}{2} + (1-\alpha)n^- \quad \text{(A52)}$$

We have the local density as

$$n_v(0) = n^+ + n^-$$

$$= \alpha\frac{n_s}{2} + (2-\alpha)n^-$$

$$= \alpha\frac{n_s}{2} + (2-\alpha)\left\{\frac{n_v(0)}{2} + \frac{1}{2}\tau\frac{dn}{dz}\left(\frac{2k_BT}{\pi m}\right)^{1/2} + \frac{3}{4} n\tau \frac{1}{T}\frac{dT}{dz}\left(\frac{2k_BT}{\pi m}\right)^{1/2}\right\} \quad \text{(A53)}$$

The above expression leads to

$$n_v(0) = n_s + \frac{(2-\alpha)}{\alpha}\tau\left(\frac{2k_BT}{\pi\, m}\right)^{1/2}\left\{\frac{dn}{dz} + \frac{3}{2}n\frac{1}{T}\frac{dT}{dz}\right\} \qquad \text{(A54)}$$

**Sec. 4. Evaporation and Condensation At A Single Interface with Diffusion Above the Interface**

**Temperature distribution.** Equation (17) is copied here

$$q = \frac{5}{2}\frac{RT}{M}\dot{m} - k_\infty\left(\frac{T}{T_\infty}\right)^{1/2}\frac{dT}{dz} \qquad \text{(A55)}$$

It is better we cast the above equation into dimensionless form by setting

$$\theta = \frac{T}{T_\infty} \qquad \text{and} \qquad \zeta = \frac{qz}{k_\infty T_\infty} \qquad \text{(A56)}$$

So that Eq. (A55) becomes

$$1 = B\theta - \theta^{1/2}\frac{d\theta}{d\zeta} \qquad \text{(A57)}$$

With

$$B = \frac{5}{2}\frac{R\dot{m}T_\infty}{Mq} \qquad \text{(A58)}$$

Equation (A57) can be written as

$$\frac{\sqrt{\theta}d\theta}{B\theta-1} = d\zeta \qquad \text{(A59)}$$

Set $t^2=\theta$, Eq. (A59) is then

$$\frac{2t^2dt}{Bt^2-1} = d\zeta \qquad \text{(A60)}$$

Integrating the above equation, we get

$$\left|\frac{\sqrt{B\theta_v(0)}-1}{\sqrt{B\theta_v(0)}+1}\right|\left|\frac{\sqrt{B\theta}+1}{\sqrt{B\theta}-1}\right| exp\left[\sqrt{B\theta_v(0)} - \sqrt{B\theta}\right] = exp\left(-\frac{B^{3/2}\zeta}{2}\right) \qquad \text{(A61)}$$

We can also obtain the temperature gradient at z=0 from the above expression

$$\left.\frac{d\theta}{d\zeta}\right]_{\zeta=0} = B\sqrt{\theta_v(0)}\frac{B\theta_v(0)-1}{B\theta_v(0)+2\sqrt{B}-1} \qquad \text{(A62)}$$

Or

$$\left.\frac{dT}{dz}\right]_{z=0} = \frac{5}{2}\frac{R\dot{m}T_\infty}{Mk_\infty}\sqrt{\frac{T_v(0)}{T_\infty}}\frac{B\frac{T_v(0)}{T_\infty}-1}{B\frac{T_v(0)}{T_\infty}+2\sqrt{B}-1} \quad \text{(A63)}$$

We can also solve Eq.(17) assuming k is a constant. In this case,

$$k\frac{dT}{dz} = \frac{5}{2}\frac{RT}{M}\dot{m} - q \quad \text{(A64)}$$

Integrating the above equation, we have

$$\left|\frac{\frac{2qM}{5R\dot{m}}-T(z)}{\frac{2qM}{5R\dot{m}}-T_v(0)}\right| = \exp\left(\frac{5}{2}\frac{R\dot{m}}{Mk}z\right) \quad \text{(A65)}$$

From which, we have

$$\left.\frac{dT}{dz}\right]_{z=0} = \left(\frac{5}{2}\frac{R\dot{m}}{Mk}\right)\left(\frac{2qM}{5R\dot{m}} - T_v(0)\right) \quad \text{(A66)}$$

**Density and Pressure Distributions.** To solve for density, $Eq.(13)$ can be written as

$$\frac{d[ln(T\rho)]}{dz} = -\frac{M}{T\rho R\tau}(\dot{m} - \rho_\infty u_\infty) \quad \text{(A67)}$$

where we have used Eq. (16). The reason we write into the above expression is since $\tau\rho \sim T^{-1/2}$, the right-hand side is actually independent of the density, and is proportional to $T^{-1/2}$, so we can integrate the above expression

$$ln\left(\frac{T(z)\rho(z)}{T_v(0)\rho_v(0)}\right) = -\int_o^z \frac{1}{\rho D}(\dot{m} - \rho_\infty u_\infty)dz \quad \text{(A68)}$$

where $D=RT\tau/M$ is the self-diffusivity, which is a function of temperature $\rho D \sim \sqrt{T}$. The above equation can be expressed as

$$ln\left(\frac{T(z)\rho(z)}{T_v(0)\rho_v(0)}\right) = ln\left(\frac{P(z)}{P_v(0)}\right) = -\frac{(\dot{m}-\rho_\infty u_\infty)}{\rho_\infty D_\infty}\int_o^z \frac{1}{\sqrt{T/T_\infty}}dz \quad \text{(A69)}$$

Using temperature distribution, the integral on the right-hand side can be performed.

**Liquid Region Temperature Distribution.** The heat flow in the liquid layer is

$$q_l = \dot{m}c_{p,l}[T(z) - T_r] - k_l\frac{dT}{dz} \quad \text{(A70)}$$

where $T_r$ is a reference temperature for the liquid enthalpy. We assume that liquid comes into the region at a temperature equaling the wall temperature. Since $q_l$ is a constant, the above equation can be easily integrated, leading to

$$\left|\frac{T(z)-T_r-\frac{q_l}{\dot{m}c_{p,l}}}{T_s-T_r-\frac{q_l}{\dot{m}c_{p,l}}}\right| = exp\left(\frac{\dot{m}c_{p,l}}{k_l}z\right)$$

We take $T_r$=$T_s$, solving for $q_l$,

$$q_l = \frac{\dot{m}c_{p,l}(T_s-T_l)}{exp\left(\frac{\dot{m}c_{p,l}}{k_l}d\right)-1} \quad \text{(A71)}$$

When $\frac{\dot{m}c_{p,l}}{k_l}\dot{d} << 1$, we can approximate the exponential by expansion, which leads to

$$q_l = \frac{k_l}{d}(T_s - T_l) \quad \text{(A72)}$$

which means that the convective term is small and we can just consider heat conduction inside the liquid layer.

**Solving for Interfacial Temperature and Density Jumps.**

Mass flux is given by

$$\dot{m} = \frac{2\alpha}{2-\alpha}\sqrt{\frac{R}{2\pi M}}\left[\rho_s(T_l)\sqrt{T_l} - \rho_v(0)\sqrt{T_v(0)}\right] \quad \text{(A73)}$$

Continuity of heat flux at liquid-vapor interface is

$$\frac{k_l}{d}(T_s - T_l) - \dot{m}L = \frac{2RT_l}{M}\dot{m} + \frac{2\alpha}{2-\alpha}\frac{R}{M}\sqrt{\frac{2R}{\pi M}}\rho_v(0)\sqrt{T_v}(T_l - T_v) = q \quad \text{(A74)}$$

Eqs. (A73), (A74) (which are two equations), together with Eqs. (A65), (A69), Eqs.(16) and (32) are a total of seven independent equations that can be used to solve for seven unknowns q, $\dot{m}$, $u_\infty$, $T_v(0)$, $\rho_v(0)$, $T_l(0), and\ T_{v,\infty}$.

**Saturation Condition.** We can use the Clausius-Clapeyron equation to relate the saturation pressure and temperature:

$$ln\frac{P_2}{P_1} = -\frac{ML}{R}\left(\frac{1}{T_2} - \frac{1}{T_1}\right) \quad \text{(A75)}$$

For water, the above expression is not accurate. I used the following correlation after fitting water $P_s(T)$ using data from the steam table,

$$\rho(T) = exp[-0.0017384T^2 + 0.1599T - 36.0032] \quad \text{(A76)}$$

Clearly, the above expression has narrow temperature range of validity.

**Sec. 5 Evaporation and Condensation at a Single Interface with Convection Above the Interface**

If above the interface, there are convection, we need to replace the diffusion equations in the vapor phase, i.e., Eqs. (13) and (14) with appropriate single phase convection boundary conditions. We copy Eqs. (13) and rewrite (14) as below:

$$\dot{m} \approx \rho u - \frac{RT}{M}\tau\frac{d\rho}{dz} - \frac{RT}{M}\rho\tau\frac{1}{T}\frac{dT}{dz} \quad \text{(A77)}$$

$$\frac{2Mq}{5RT} \approx \rho u - \tau\frac{RT}{M}\frac{d\rho}{dz} - 2\frac{RT}{M}\rho\tau\frac{1}{T}\frac{dT}{dz} \quad \text{(A78)}$$

Subtracting the two equations, we get

$$\frac{RT}{M}\rho\tau\frac{1}{T}\frac{dT}{dz} = \dot{m} - \frac{2Mq}{5RT} \quad \text{(A79)}$$

Substituting (A79) into (A77), we get

$$\frac{RT}{M}\tau\frac{d\rho}{dz} = \rho u - 2\dot{m} + \frac{2Mq}{5RT} \quad \text{(A80)}$$

Substituting Eqs. (A79) and (A80) into Eq. (31) leads to

$$\rho_v(0) = \rho_s + \frac{(2-\alpha)}{\alpha}\left(\frac{2M}{\pi RT}\right)^{1/2}\left\{\rho u - \frac{\dot{m}}{2} - \frac{Mq}{5RT}\right\} \quad \text{(A81)}$$

Replacing $\dot{m}$ and q in the above equation with Eqs. (42) and (43), we obtain

$$\rho_v(0) = \rho_s - \frac{(2-\alpha)}{\alpha}\left(\frac{2M}{\pi RT_v}\right)^{1/2}\left[\frac{h_{s,v}}{\rho c_p}(\rho_v - \rho_\infty) + \frac{M}{5RT_v(0)}h_{s,v}[T_v(0) - T_\infty]\right] \quad \text{(A82)}$$

There is the question what density we should use in the denominator of first term in the square bracket. Based on the fact that for heat flux, we have $T_v(0)$ in the denominator, it is probably best if we use the density as $\rho_v$. There is also the choice of taking an average or using $\rho_\infty$. The differences among the choices are likely to be small and our solution used $\rho_\infty$.